\documentclass[twocolumn,floats,floatfix,aps,pra]{revtex4}
\usepackage{amsfonts,amssymb,amsmath}
\usepackage{color,calc}
\usepackage[dvips]{graphicx}
\usepackage{bm}
\usepackage{amsbsy,amsmath}
\usepackage{array}
\usepackage{booktabs}

\def\be{ \begin{equation} }
\def\ee{ \end{equation} }
\def\bea{ \begin{eqnarray} }
\def\eea{ \end{eqnarray} }
\def\bse{ \begin{subequations} }
\def\ese{ \end{subequations} }
\def\ba{ \begin{array} }
\def\ea{ \end{array} }
\def\bt{ \begin{tabular} }
\def\et{ \end{tabular} }

\def\i{\,\text{i}}
\def\e{\,\text{e}}
\def\i{i}
\def\e{e}

\def\to{\rightarrow}

\def\d{\text{d}}

\def\U{\mathbf{U}}
\def\F{\mathbf{F}}
\def\H{\mathbf{H}}

\def\A{\mathcal{A}}

\newcommand{\ket}[1]{\vert #1\rangle}
\def\i{{\rm{i}}}
\def\f{{\rm{f}}}
\def\phase{\phi}

\begin{document}

\title{Ultrahigh-fidelity composite rotational quantum gates}

\author{Hayk Gevorgyan}
\affiliation{Faculty of Physics, St Kliment Ohridski University of Sofia, 5 James Bourchier blvd, 1164 Sofia, Bulgaria}

\author{Nikolay V. Vitanov}
\affiliation{Faculty of Physics, St Kliment Ohridski University of Sofia, 5 James Bourchier blvd, 1164 Sofia, Bulgaria}

\date{\today }

\begin{abstract}
Composite pulse sequences, which produce arbitrary pre-defined rotations of a qubit on the Bloch sphere, are presented.
The composite sequences contain up to 17 pulses and can compensate up to eight orders of experimental errors in the pulse amplitude and the pulse duration.
Composite sequences for three basic quantum gates --- X (NOT), Hadamard and arbitrary rotation --- are derived.
Three classes of composite sequences are presented --- one symmetric and two asymmetric.
They contain as their lowest members two well-known composite sequences --- the three-pulse symmetric SCROFULOUS pulse and the four-pulse asymmetric BB1 pulse, which compensate first and second-order errors, respectively.
The shorter sequences are derived analytically, and the longer ones numerically (instead by nesting and concatenation, as mostly done hitherto).
Consequently, the composite sequences derived here match or outperform the existing ones in terms of either speed or accuracy, or both.
For example, we derive a second-order composite sequence, which is faster (by about 13\%) than the famous BB1 sequence.
For higher-order sequences the speed-up becomes much more pronounced.
This is important for quantum information processing as the sequences derived here provide more options for finding the sweet spot between ultrahigh fidelity and high speed.
\end{abstract}

\maketitle

%\blue

%%%%%%%%%%%%%%%%%%%%%%%%%%%%%%%%%%%%%%%%%%%%%%%%%%%%%%%%%%%%%%%%%%%%%%%%%%%%%%%%%%%%%%%%%%%%%%%%%%%%%%%%%%%%%%%%%%%%%%%%%%%%%%%%%%%%%%%%%
\section{Introduction\label{Sec:intro}}

Quantum rotation gates, such as the Hadamard gate and the X (or NOT) gate are central elements in any quantum circuit \cite{Nielsen2000,Vandersypen2004,Jones2011}.
Traditionally, a general rotation at an angle $\theta$ is implemented by a resonant pulsed field with a temporal area of $\theta$, hence the name $\theta$ pulses.
In particular, the Hadamard gate is implemented by a resonant $\pi/2$ pulse, and the X gate is implemented by a resonant $\pi$ pulse, which are the theoretically fastest means for producing these gates.
However, resonant driving is prone to errors in the experimental parameters, e.g. the pulse amplitude, duration, and detuning.

Various proposals have been made in order to generate rotation gates that are resilient to experimental errors, at the expense of being longer, and hence slower.
Adiabatic techniques are the traditional remedy for tackling such errors \cite{Vitanov2001}.
Ever since 1932 \cite{Landau1932,Majorana1932,Stuckelberg1932,Zener1932}, adiabatic evolution via a level crossing is the ubiquitous method to produce complete population inversion and hence the X gate.
More recently, adiabatic evolution via a half crossing has gained popularity as a means for producing half excitation, and hence the Hadamard gate  \cite{Yatsenko2002,Vitanov2006,Yamazaki2008,Zlatanov2017,Randall2018}.
This idea has been used in a technique known as half-SCRAP (Stark-chirped rapid adiabatic passage) \cite{Yatsenko2002} and the closely related two-state STIRAP (stimulated Raman adiabatic passage) \cite{Vitanov2006}, which has been successfully implemented in a trapped-ion experiment \cite{Yamazaki2008}.
In both cases, pulse shaping and chirping are designed such that their time dependences resemble the delayed-pulse ordering of conventional STIRAP \cite{Vitanov2017}.
In a variation of these, an adiabatic technique has been proposed \cite{Zlatanov2017} which generates arbitrary coherent superpositions of two states, which is controlled by the initial and final ratios of the field's amplitude and its detuning.
An extension of this half-crossing technique to three states has been experimentally demonstrated in a trapped-ion experiment, with an error of about $1.4 \times 10^{-4}$, i.e. close to the quantum computation benchmark level \cite{Randall2018}, which was achieved by using pulse shaping.
Another proposal used a sequence of two half-crossing adiabatic pulses split by a phase jump, which serves as a control parameter to the created superposition state \cite{Zlatanov2020}.

In three-state Raman-coupled qubits, a very popular technique is fractional STIRAP \cite{Marte1991,Weitz1994,Vitanov1999}, in which the Stokes pulse arrives before the pump pulse but the two pulses vanish simultaneously.
This leads to the creation of a coherent superposition of the two end states of the chain.
Tripod-STIRAP \cite{Unanyan1998,Theuer1999,Vewinger2003}, an extension of STIRAP wherein a single state is coupled to three other states, has also been used for the generation of coherent superpositions of these three states or two of them.
We also note a technique for creation of coherent superposition states and for navigation between them by quantum Householder reflections \cite{Ivanov2007,Rousseaux2013}.

While adiabatic techniques provide great robustness to parameter errors, in general they struggle to deliver the ultrahigh fidelity required in quantum computation.
A powerful alternative to achieve ultrahigh fidelity while featuring robustness to parameter errors is the technique of composite pulses \cite{Levitt1979,Freeman1980,Levitt1982,Levitt1983,Levitt1986}.
The composite pulse sequence is a finite train of pulses with well-defined relative phases between them.
These phases are control parameters, which are determined by the desired excitation profile.
Composite pulses can shape the excitation profile in essentially any desired manner, which is impossible with a single resonant pulse or adiabatic techniques.
In particular, one can create a broadband composite $\pi$ pulse, which delivers transition probability of 1 not only for a pulse area $\A=\pi$ and zero detuning $\Delta=0$, as a single resonant $\pi$ pulse, but also in some ranges around these values \cite{Levitt1979,Freeman1980,Levitt1982,Levitt1983,Levitt1986,Wimperis1990,Wimperis1991,Wimperis1994,Levitt2007,Torosov2011PRA,Torosov2011PRL,Schraft2013,Genov2014}.
%Thus a composite pulse can compensate the imperfections of a single real $\pi$ pulse and make it look like an ideal $\pi$ pulse.
Alternatively, narrowband composite pulses \cite{Tycko1984,Tycko1985,Shaka1984,Wimperis1990,Wimperis1994,Torosov2011PRA, Vitanov2011,Ivanov2011,Merrill2014} squeeze the excitation profile around a certain point in the parameter space: they produce excitation that is more sensitive to parameter variations than a single pulse, with interesting applications to sensing, metrology and spatial localization in NMR spectroscopy.
A third family of composite pulses --- passband pulses --- combine the features of broadband and narrowband pulses: they provide highly accurate excitation inside a certain parameter range and negligibly small excitation outside it \cite{Cho1986,Cho1987,Wimperis1989, Wimperis1994, Ivanov2011, Kyoseva2013}.

There are no universally applicable composite pulses because the requirements in different applications are different.
For instance, in NMR, composite pulses which compensate errors in very broad parameter ranges with only modest accuracy are ubiquitous.
On the contrary, in quantum information, very high accuracy is required within some moderately large parameter ranges \cite{Gulde2003,Schmidt-Kaler2003,Haffner2008,Timoney2008,Monz2009,Zarantonello2019}.

In this paper, we present several sets of single-qubit rotation quantum gates constructed with composite pulse sequences.
There are two classes of composite rotations, named variable and constant rotations \cite{Levitt1986,Levitt2007}.
\emph{Variable-rotation} composite pulses (sometimes called \emph{Class B}) compensate parameter errors only in the transition probability $p$ (or the population inversion $w = 2p - 1$).
Recently \cite{Torosov2019variable}, several classes of arbitrarily accurate analytic composite sequences for variable rotations have been presented.
\emph{Constant-rotation}, or \emph{phase-distortionless} \cite{Tycko1985jmr}, composite pulses (sometimes called \emph{Class A}) compensate parameter errors in both the transition probability and the phases of the created superposition state (i.e., in the Bloch vector coherences $u$ and $v$).
The latter are obviously more demanding and require longer sequences for the same order of compensation.
However, in quantum information processing wherein phase relations are essential, constant rotations are clearly the ones to be used for quantum rotation gates \cite{Cummins2003}.

In this paper, we focus at the derivation of ultrahigh-fidelity composite rotation gates, including the X, Hadamard and general rotation, which compensate pulse-area errors up to eighth order.
The X and Hadamard gates are special cases of general rotations but they are treated separately due to their importance in quantum information.
Our results extend earlier results on some of these gates using shorter pulse sequences.
The first phase-distortionless composite pulse was designed by Tycko \cite{Tycko1984} which produces a composite X gate.
It consists of three pulses of total nominal area of $3\pi$ and provides a first-order error compensation.
A second-order error compensation composite pulse was constructed by Wimperis, the well-known BB1 pulse \cite{Wimperis1991,Wimperis1994}.
It consists of four pulses with a total nominal pulse area of $4\pi+\theta$ and it produces a constant rotation at an arbitrary angle $\theta$.
More recently, Wimperis and co-workers developed several phase-distortionless anti-symmetric composite $\pi$ pulses designed for rephasing of coherence \cite{Odedra2012a,Odedra2012b,Odedra2012c}.
Jones and co-workers have devoted a great deal of attention to composite X gates, with an emphasis of geometric approaches for derivation of such sequences, which work up to 5 and 7 pulses \cite{Cummins2003,Jones2013pra,Jones2013pla,Husain2013}.

Composite rotation gates with a pulse area error compensation of third and higher order have been constructed using nesting and concatenation of shorter composite sequences.
For larger error order, this procedure produces (impractical) composite sequences of extreme length.
Here we use analytic approaches and brute-force numerics to derive three classes of composite sequences for X, Hadamard and rotation gates which achieve error compensation of up to 8th order with much shorter sequences than before.

This paper is organized as follows.
In Sec.~\ref{Sec:derivation} we explain the derivation method.
Composite $\pi$ rotations, representing the X gate are presented in Sec.~\ref{Sec:X}.
Composite implementations of the Hadamard gate are given in Sec.~\ref{Sec:H}, and composite rotation gates in Sec.~\ref{Sec:rotation}.
Finally, Sec.~\ref{Sec:conclusion} presents the conclusions.

%%%%%%%%%%%%%%%%%%%%%%%%%%%%%%%%%%%%%%%%%%%%%%%%%%%%%%%%%%%%%%%%%%%%%%%%%%%%%%%%%%%%%%%%%%%%%%%%%%%%%%%%%%%%%%%%%%%%%%%%%%%%%%%%%%%%%%%%%
%%%%%%%%%%%%%%%%%%%%%%%%%%%%%%%%%%%%%%%%%%%%%%%%%%%%%%%%%%%%%%%%%%%%%%%%%%%%%%%%%%%%%%%%%%%%%%%%%%%%%%%%%%%%%%%%%%%%%%%%%%%%%%%%%%%%%%%%%
\section{Composite rotation gates: derivation \label{Sec:derivation}}
%%%%%%%%%%%%%%%%%%%%%%%%%%%%%%%%%%%%%%%%%%%%%%%%%%%%%%%%%%%%%%%%%%%%%%%%%%%%%%%%%%%%%%%%%%%%%%%%%%%%%%%%%%%%%%%%%%%%%%%%%%%%%%%%%%%%%%%%%
%%%%%%%%%%%%%%%%%%%%%%%%%%%%%%%%%%%%%%%%%%%%%%%%%%%%%%%%%%%%%%%%%%%%%%%%%%%%%%%%%%%%%%%%%%%%%%%%%%%%%%%%%%%%%%%%%%%%%%%%%%%%%%%%%%%%%%%%%
\def\R{\hat{R}}
\def\F{\hat{F}}
\def\I{\mathbf{I}}

\subsection{Composite rotation gates}

Our objective is to construct the qubit rotation gate $\R_y(\theta) = e^{i (\theta/2) \hat\sigma_y}$, where $\theta$ is the rotation angle and $\hat\sigma_y$ is the Pauli's $y$ matrix.
In matrix form,
\be\label{rot-gate}
\mathbf{R}_y(\theta) =\left[ \begin{array}{cc} \cos (\theta/2) & \sin (\theta/2) \\  -\sin (\theta/2) & \cos (\theta/2) \end{array} \right].
\ee
The rotation gate \eqref{rot-gate} is equivalent to the rotation gate $\R_x(\theta) = e^{i (\theta/2) \hat\sigma_x}$, or in matrix form,
\be\label{rot-gate-x}
\mathbf{R}_x(\theta) = \left[ \begin{array}{cc} \cos (\theta/2) & i \sin (\theta/2) \\  i \sin (\theta/2) & \cos (\theta/2) \end{array} \right].
\ee
Indeed, $\R_x(\theta)$ can be obtained from  $\R_y(\theta)$ by simple phase transformation,  $\R_x(\theta) = \F(\pi/4) \R_y(\theta) \F(-\pi/4)$. Here $\F(\phi) = e^{i \phi \hat\sigma_z}$, or in matrix form,
\be\label{F}
\mathbf{F}(\phi) = \mathbf{R}_z(\phi) = \left[ \begin{array}{cc}  e^{i\phi} & 0 \\  0 & e^{-i\phi} \end{array} \right].
\ee
We shall use the gate \eqref{rot-gate} because it is real and because it coincides with the ubiquitous definition of the rotation matrix.
Therefore, hereafter we drop the subscript $y$ for the sake of brevity.

The propagator of a coherently driven qubit is the solution of the Schr\"odinger equation,
\be
i \hbar \partial_t \U(t,t_i) = \H(t) \U(t,t_i),
\ee
subject to the initial condition $\U(t_i,t_i) = \I$, the identity matrix.
If the Hamiltonian is Hermitian, the propagator is unitary.
If the Hamiltonian is also traceless, then the propagator has the SU(2) symmetry and can be represented as
\be\label{SU(2)}
\U_0 = \left[ \begin{array}{cc} a & b \\ -b^{\ast} & a^{\ast} \end{array}\right],
\ee
where $a$ and $b$ are the complex-valued Cayley-Klein parameters satisfying $|a|^2+|b|^2=1$.
A traceless Hermitian Hamiltonian has the form
$\hat{H}(t) = \frac12 \hbar [ \Omega(t) \cos(\phi) \hat{\sigma}_x + \Omega(t) \sin(\phi) \hat{\sigma}_y + \Delta \hat{\sigma}_z]$, where $\Omega(t)$ (assumed real and positive) is the Rabi frequency quantifying the coupling, $\phi$ is its phase, and $\Delta$ is the field-system detuning.

On exact resonance ($\Delta=0$) and for $\phi=0$, we have $a=\cos(\A/2) $, $b=-\i\sin(\A/2)$, where $\A$ is the temporal pulse area $\A=\int_{t_\i}^{t_\f}\Omega(t)\d t$.
For a system starting in state $\ket{1}$, the single-pulse transition probability is $p = |b|^2=\sin^2 (\A/2)$.

A single resonant pulse of temporal area $\A=\theta_\epsilon = \theta (1+\epsilon)$ produces the propagator $\R(\theta_\epsilon) = e^{i [\theta (1+\epsilon)/2] \hat\sigma_y} = \R(\theta) [1 + O(\epsilon)]$, i.e. it is accurate up to zeroth order $O(\epsilon^0)$ in the pulse area error $\epsilon$.
Our approach is to replace the single $\theta$ pulse with a composite sequence of pulses of appropriate pulse areas and phases, such that the overall propagator produces the rotation gate \eqref{rot-gate} with an error of higher order, i.e. $\R(\theta) [1 + O(\epsilon^{n+1})]$. Then we say that the corresponding composite rotation gate is accurate up to, and including, order $O(\epsilon^{n})$.

\subsection{Derivation}

The derivation of the composite rotation gates is done in the following manner.
A phase shift $\phase$ imposed on the driving field, $\Omega(t)\to\Omega(t)\e^{\i\phase}$, is imprinted onto the propagator \eqref{SU(2)} as
\be\label{U phase}
\U_\phase = \left[ \begin{array}{cc} a & b \e^{\i\phase} \\ -b^{\ast}\e^{-\i\phase} & a^{\ast} \end{array}\right].
\ee
A train of $N$ pulses, each with area $\A_k$ and phase $\phase_k$ (applied from left to right),
\be
(\A_1)_{\phi_1} (\A_2)_{\phi_2} (\A_3)_{\phi_3} \cdots (\A_N)_{\phi_{N}},
\ee
produces the propagator (acting, as usual, from right to left)
\be\label{U^N}
\boldsymbol{\mathcal{U}} = \U_{\phase_{N}}(\A_N) \cdots \U_{\phase_{3}}(\A_3) \U_{\phase_{2}}(\A_2) \U_{\phase_{1}}(\A_1).
\ee
Let us assume that the nominal (i.e. for zero error) pulse areas $A_k$ have a systematic error $\epsilon$, i.e. $A_k \to A_k (1+\epsilon)$.
If all nominal pulse areas are the same, as it is the case for many composite sequences, this is the natural assumption because the apparatus will produce possibly imperfect but identical pulses.
If the pulse areas are different, this is also a reasonable assumption in many cases.
For example, if a trapped ion is addressed by an imperfectly pointed laser beam then it will ``see'' the same systematic deviation from the perfect field amplitude (and hence pulse area) for any chosen target pulse area.
Atoms in atomic clouds in magnetooptical or dipole traps or ions in doped solids (e.g. for optical memories) addressed by electromagnetic fields offer another example: they will ``see'' different field amplitude due to spatial inhomogeneity depending on their position in the sample, but this field amplitude will deviate from the optimal one by the same relative systematic error $\epsilon$ regardless of the value of the optimal amplitude if the atoms do not move much during the duration of the composite sequence.

Under the assumption of a single systematic pulse area error $\epsilon$, we can expand the composite propagator \eqref{U^N} in a Taylor series versus $\epsilon$.
Because of the SU(2) symmetry of the overall propagator, it suffices to expand only two of its elements, say $\mathcal{U}_{11}(\epsilon)$ and $\mathcal{U}_{12}(\epsilon)$.
We set their zero-error values to the target values,
\be\label{eq-0}
\mathcal{U}_{11}(0) = \cos(\theta/2),\quad \mathcal{U}_{12}(0) = \sin(\theta/2),
\ee
and we set as many of their derivatives with respect to $\epsilon$, in the increasing order, as possible,
\be\label{eq-m}
\mathcal{U}^{(m)}_{11}(0) = 0,\quad \mathcal{U}^{(m)}_{12}(0) = 0, \quad (m=1,2,\ldots, n),
\ee
where $ \mathcal{U}^{(m)}_{jl} = \partial_\epsilon^m  \mathcal{U}_{jl}$ denotes the $m$th derivative of $\mathcal{U}_{jl}$ with respect to $\epsilon$.
The largest derivative order $n$ satisfying Eqs.~\eqref{eq-m} gives the order of the error compensation $O(\epsilon^n)$.

Equations \eqref{eq-0} and \eqref{eq-m} generate a system of $2(n+1)$ algebraic equations for the nominal pulse areas $A_k$ and the composite phases $\phi_k$ ($k=1,2,\ldots,N$).
The equations are complex-valued and generally we have to solve $4(n+1)$ equations with the $2N$ free parameters (nominal pulse areas and phases).
Because of the normalization condition $|\mathcal{U}_{11}|^2 + |\mathcal{U}_{12}|^2 = 1$, an error compensation of order $n$ requires a composite sequence of  $N=2n+1$ pulses (or $N=2n$ in some lucky cases).

As stated above, the derivation of the composite sequences requires the solution of Eqs.~\eqref{eq-0} and \eqref{eq-m}.
For a small number of pulses (up to about five), the set of equations can be solved analytically.
For longer sequences, Eqs.~\eqref{eq-0} and the first two equations ($n=1$) of Eqs.~\eqref{eq-m} can still be solved analytically, but the higher orders in Eqs.~\eqref{eq-m} they are solved numerically.
We do this by using standard routines in \textsc{Mathematica}%\texttt{Mathematica}
$^\copyright$.

\subsection{Quantum gate fidelity}

If Eqs.~\eqref{eq-0} and \eqref{eq-m} are satisfied, then the overall propagator can be written as
\be\label{U-epsilon}
\boldsymbol{\mathcal{U}} (\epsilon) = \mathbf{R}(\theta) + O(\epsilon^{n+1}),
\ee
with $\mathbf{R}(\theta) = \boldsymbol{\mathcal{U}} (0)$.
Then the \emph{Frobenius distance fidelity},
\be\label{Frobenius}
\mathcal{F} = 1 - \| \boldsymbol{\mathcal{U}} (\epsilon) - \mathbf{R}(\theta) \|
 = 1 - \sqrt{ \tfrac14 \sum\nolimits_{j,k=1}^2 \left|\mathcal{U}_{jk} - R_{jk} \right|^2 } ,
\ee
is of the same error order $O(\epsilon^{n})$ as the propagator, $\mathcal{F} = 1 - O(\epsilon^{n+1})$.
As shown by Jones and co-workers \cite{Jones2011} for the composite X gates, the \emph{trace fidelity},
\be\label{trace fidelity}
\mathcal{F}_{\text{T}} = \tfrac12 \text{Tr}\, [ \boldsymbol{\mathcal{U}} (\epsilon) \mathbf{R}(\theta)^\dagger ] ,
\ee
has a factor of 2 higher error order $O(\epsilon^{2n})$, i.e. $\mathcal{F}_{\text{T}} = 1 - O(\epsilon^{2n+1})$.
The reason is that in the Frobenius distance, all information of the actual propagator is involved, while in the trace distance some of this information is lost.
Therefore, throughout this paper we shall use the Frobenius distance fidelity \eqref{Frobenius}, which is a much more strict and unforgiving to errors fidelity measure; moreover, its error is of the same order as the propagator error.

We note here that for variable rotations, Eqs.~\eqref{eq-0} and \eqref{eq-m} have to be satisfied for only one of the propagator elements, say $\mathcal{U}_{12}$.
This means that with the same number of pulses one can achieve a factor of 2 higher order of error compensation for variable rotations than for constant rotations.
However, this error compensation applies to the transition probability only, but not to the propagator phases.
For variable rotations the overall propagator cannot be written in the form of Eq.~\eqref{U-epsilon}, and consequently, neither of the fidelities \eqref{Frobenius} or \eqref{trace fidelity} is of the form $ 1 - O(\epsilon^{n+1})$.

\subsection{Composite pulse sequences}

Based on numerical evidence, we consider three types of composite sequences, one symmetric and two asymmetric.

\begin{itemize}

\item
Each symmetric sequence consists of a sequence of $2n-1$ nominal $\pi$ pulses, sandwiched by two pulses of areas $\alpha$, with symmetrically ordered phases,
\be\label{CP-symmetric}
\alpha_{\phi_1} \pi_{\phi_2} \pi_{\phi_3} \cdots \pi_{\phi_{n-1}} \pi_{\phi_n} \pi_{\phi_{n-1}} \cdots \pi_{\phi_3} \pi_{\phi_2} \alpha_{\phi_1}.
\ee
These sequences generalize the three-pulse SCROFULOUS sequence \cite{Cummins2003}, which is of this type, to more than three pulses.

\item
The first type of asymmetric sequences consists of a sequence of nominal $\pi$ pulses, preceded (or superseded) by a pulse of area $\theta$,
\be\label{CP-asymmetric-theta}
\pi_{\phi_1} \pi_{\phi_2} \pi_{\phi_3} \cdots \pi_{\phi_{N-1}} \theta_{\phi_N} .
\ee
These sequences generalize the five-pulse BB1 sequence \cite{Wimperis1994}, which is of this type, to more than five pulses.

\item
The second type of asymmetric sequences consists of a sequence of $N-2$ nominal $\pi$ pulses, preceded (or superseded) by single pulses of areas $\alpha$ and $\beta$,
\be\label{CP-asymmetric-ab}
\alpha_{\phi_1} \pi_{\phi_2} \pi_{\phi_3} \cdots \pi_{\phi_{N-1}} \beta_{\phi_N} .
\ee
To the best of our knowledge, this type of composite sequences has not been reported in the literature hitherto.

\end{itemize}

Below we consider these three classes of composite sequences and test their performance by using the Frobenius distance \eqref{Frobenius}.
We consider three figures of merit to be essential.

\begin{itemize}

\item The most important parameter is the order of error compensation $O(\epsilon^n)$.
The larger $n$, the broader the high-fidelity range and the larger the errors $\epsilon$, which can be compensated.

\item The second most important parameter is the total pulse area $\mathcal{A}_{\text{tot}} = \sum_{k=1}^N |\mathcal{A}_k|$.
It determines the length of the sequences and hence the speed of the gates.
Usually, the peak Rabi frequency is limited either by the experimental apparatus or by the qubit properties, e.g., too large Rabi frequency can cause unwanted couplings to other levels or to other qubits (cross-talk).
Therefore, for a fixed peak Rabi frequency, the total pulse area determines the total duration of the composite sequence.

\item Another consideration is the number of pulses $N$ in the sequence.
Unless there are issues with the implementation of the phase jumps, this argument is of far less importance than the other two.
However, if the phase jumps require some time to implement or cannot be implemented with high accuracy, then sequences of fewer pulses are preferable.
For this reason, we often give several different CPs for each error order.

\end{itemize}

%%%%%%%%%%%%%%%%%%%%%%%%%%%%%%%%%%%%%%%%%%%%%%%%%%%%%%%%%%%%%%%%%%%%%%%%%%%%%%%%%%%%%%%%%%%%%%%%%%%%%%%%%%%%%%%%%%%%%%%%%%%%%%%%%%%%%%%%%
%%%%%%%%%%%%%%%%%%%%%%%%%%%%%%%%%%%%%%%%%%%%%%%%%%%%%%%%%%%%%%%%%%%%%%%%%%%%%%%%%%%%%%%%%%%%%%%%%%%%%%%%%%%%%%%%%%%%%%%%%%%%%%%%%%%%%%%%%
%\section{Composite rotation gates: symmetric sequences \label{Sec:rot-symmetric}}
%%%%%%%%%%%%%%%%%%%%%%%%%%%%%%%%%%%%%%%%%%%%%%%%%%%%%%%%%%%%%%%%%%%%%%%%%%%%%%%%%%%%%%%%%%%%%%%%%%%%%%%%%%%%%%%%%%%%%%%%%%%%%%%%%%%%%%%%%
%%%%%%%%%%%%%%%%%%%%%%%%%%%%%%%%%%%%%%%%%%%%%%%%%%%%%%%%%%%%%%%%%%%%%%%%%%%%%%%%%%%%%%%%%%%%%%%%%%%%%%%%%%%%%%%%%%%%%%%%%%%%%%%%%%%%%%%%%

\section{X (NOT) gate}\label{Sec:X}

The X or NOT gate is defined as
\be\label{NOT}
  \left[ \begin{array}{cc} 0 & 1 \\ 1 & 0 \end{array}\right] = \hat{\sigma}_x,
\ee
Because the determinant of this matrix is $-1$, it is not of SU(2) type. Instead, we shall construct the SU(2) gate
\be\label{X}
\mathbf{X} = \left[ \begin{array}{cc} 0 & 1 \\ -1 & 0 \end{array}\right],
\ee
which is related to the gate \eqref{NOT} by a phase transformation and it is equivalent to it.
The gate \eqref{X} is also equivalent to the often used gate
\be\label{e(ix)}
e^{i (\pi/2) \hat{\sigma}_x} = \left[ \begin{array}{cc} 0 & i \\ i & 0 \end{array}\right],
\ee
which can be obtained from Eq.~\eqref{X} by a phase transformation too.
However, we prefer to use the gate \eqref{X} because it is real and also because it is a special case of the general rotation gate \eqref{rot-gate}.

As it is well known, such a gate can be produced by a resonant pulse of temporal area $\pi$.
The propagator of a $\pi$ pulse reads
\be\label{U-pi}
\U = \left[ \begin{array}{cc} \cos(\pi(1+\epsilon)/2) & \sin(\pi(1+\epsilon)/2) \\ -\sin(\pi(1+\epsilon)/2) & \cos(\pi(1+\epsilon)/2) \end{array}\right],
\ee
where $\epsilon$ is the pulse area error.
The Frobenius distance fidelity \eqref{Frobenius} reads
\be \label{F-1-distance}
\mathcal{F} = 1 - \sqrt{2} \left| \sin \frac{\pi  \epsilon }{4} \right|.
\ee
For comparison, the trace fidelity is
\be \label{F-1-trace}
\mathcal{F}_T = 1 - 2  \sin^2 \frac{\pi  \epsilon }{4} = \cos \frac{\pi  \epsilon }{2}.
\ee
Obviously the error stemming from the Frobenius distance fidelity \eqref{F-1-distance}, which is of order $O(\epsilon)$, is far greater than the value of the error stemming from the trace fidelity \eqref{F-1-trace}, which is of order $O(\epsilon^2)$, as noted by Jones and co-workers \cite{Cummins2003}.

The three types of composite sequences \eqref{CP-symmetric}, \eqref{CP-asymmetric-theta}, and \eqref{CP-asymmetric-ab} coalesce into a single type, a sequence of $\pi$ pulses.
Below we consider these sequences, in the increasing order of error compensation.

\subsection{First-order error compensation}\label{Sec:X-3}

The careful analysis of Eqs.~\eqref{eq-0} and \eqref{eq-m} shows that the shortest possible CP which can compensate first-order errors consists of three pulses, each with a pulse area of $\pi$, and symmetric phases,
\be\label{X3}
\pi_{\phi_1} \pi_{\phi_2} \pi_{\phi_1}.
\ee
Solving Eq.~\eqref{eq-0} along with Eq.~\eqref{eq-m} for the first derivatives gives two solutions for the phases,
\bse\label{X3-phases}
\begin{align}
& \pi_{\frac16\pi} \pi_{\frac56\pi} \pi_{\frac16\pi}, \\
& \pi_{\frac56\pi} \pi_{\frac16\pi} \pi_{\frac56\pi}.
\end{align}
\ese
These two sequences generate the same propagator and hence the same fidelity.

The Frobenius distance and trace distance fidelities read
\bse
\begin{align}
& \mathcal{F} = 1 - \mathcal{I}_1 ,  \label{F-3-distance} \\
& \mathcal{F}_T = 1 - \mathcal{I}_1^2 ,  \label{F-3-trace}
%& \mathcal{F} = 1 - \sqrt{2}  \sqrt{2 + \cos \frac{\pi  \epsilon }{2}}\, \sin ^2 \frac{\pi  \epsilon }{4} ,  \label{F-3-distance} \\
%& \mathcal{F}_T = 1 - 2 \left(2 + \cos \frac{\pi  \epsilon }{2} \right) \sin ^4 \frac{\pi  \epsilon }{4} .  \label{F-3-trace}
\end{align}
\ese
where the Frobenius distance infidelity is
\be
\mathcal{I}_1 = \sqrt{2 \left(1 + 2\cos^2 \frac{\pi  \epsilon }{4}\right) }\, \sin ^2 \frac{\pi  \epsilon }{4}.
\ee
Obviously, the  Frobenius distance infidelity $\mathcal{I}_1$ is of order $O(\epsilon^2)$ and it is much larger than the trace distance infidelity $\mathcal{I}_1^2$, which is of order $O(\epsilon^4)$.

%***************************************************************
\begin{figure}[t]
\bt{c}
\includegraphics[width=0.90\columnwidth]{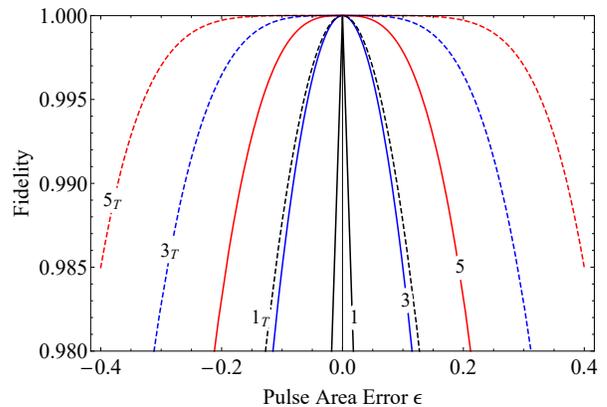}
\et
\caption{
Frobenius distance fidelity $\mathcal{F}$ (solid) and trace distance fidelity $\mathcal{F}_T$ (dashed) of composite X gates.
The numbers $N$ on the curves refer to composite sequences X$N$ listed in Table \ref{Table:X}.
}
\label{fig:X135}
\end{figure}
%***************************************************************

The Frobenius distance fidelity and the trace fidelity are plotted in Figure \ref{fig:X135} for X gates produced by a single pulse and composite sequences of 3 and 5 (see below) pulses.
The three-pulse composite X gate \eqref{X3-phases} produces much higher fidelity that the single-pulse X gate.
Obviously, the trace distance fidelity is much higher than the Frobenius distance fidelity: compare the curves with labels 1 and $1_T$; 3 and $3_T$;  5 and $5_T$.
In fact, as seen in the figure, the trace distance fidelity for a single pulse (label $1_T$) almost coincides with the Frobenius distance fidelity for the three-pulse composite sequence (label 3).
With respect to the quantum computation benchmark fidelity value of $1-10^{-4}$, the Frobenius distance fidelity \eqref{F-3-distance} for the three-pulse composite X gates of Eqs.~\eqref{X3-phases} remains above this value in the pulse area interval $(0.992\pi, 1.008\pi)$, i.e. for relative errors up to $|\epsilon| < 0.008$.
For comparison, the trace distance fidelity \eqref{F-3-trace} remains above this value in the pulse area interval $(0.919\pi, 1.081\pi)$, i.e. for relative errors up to $|\epsilon| < 0.081$, a factor of 10 larger.
This is the reason why in this work, we will use the much more severe Frobenius distance fidelity. % \eqref{F-3-distance}.

\subsection{Second-order error compensation}\label{Sec:X-4}

For sequences of four pulses, it becomes possible to annul the second-order derivatives in Eq.~\eqref{eq-m}.
A number of solutions exist, some of which are
\bse
\begin{align}
& (2\pi)_{3\chi} \pi_{\pi+\chi} \pi_{\frac12 \pi} \pi_{-\chi}, \\
& \pi_{\pi+\chi} (2\pi)_{3\chi} \pi_{\pi+\chi} \pi_{\frac12 \pi}, \\
& \pi_{\frac12 \pi} \pi_{\pi+\chi} (2\pi)_{3\chi} \pi_{\pi+\chi}, \\
& \pi_{-\chi} \pi_{\frac12 \pi} \pi_{\pi+\chi} (2\pi)_{3\chi},
\end{align}
\ese
where $\chi = \arcsin(\frac14) \approx 0.0804\pi$.
The second and third sequences are related to the BB1 sequence of Wimperis \cite{Wimperis1994}.
Note that all these sequences have a total nominal pulse area of $5\pi$, and can be considered as five-pulse sequences because the effect of $(2\pi)_{3\chi}$ is the same as $\pi_{3\chi}\pi_{3\chi}$.

The Frobenius fidelity for all these sequences reads $\mathcal{F} = 1 - \mathcal{I}_2 $, with the infidelity
\be\label{F-X-4}
\mathcal{I}_2 =  \sqrt{ 8 + 9 \cos \frac{\pi  \epsilon }{2} + 3 \cos^2 \frac{\pi  \epsilon }{2}}\,
\left|\sin \frac{\pi  \epsilon }{4}\right|^3.
\ee
Obviously, this fidelity is accurate up to order $O(\epsilon^2)$, as the error is of order $O(\epsilon^3)$.
The trace fidelity reads $\mathcal{F}_T = 1 - \mathcal{I}_2^2$.
The trace fidelity is accurate up to order $O(\epsilon^5)$, as the error is of order $O(\epsilon^6)$.
Obviously, the trace infidelity is much smaller than the Frobenius distance infidelity, as for the three-pulse composite sequences.

%T%T%T%T%T%T%T%T%T%T%T%T%T%T%T%T%T%T%T%T%T%T%T%T%T%T%T%T%T%T%T%T%T%T%T%T%T%T%T%T%T%T%T%T%T
\begin{table*}
\begin{tabular}{|c|c|c|l|c|}
\hline
Name & Pulses & $O(\epsilon^n)$ & Phases $\phi_1,\phi_2, \ldots, \phi_n$ (in units $\pi$) & High-fidelity error \\
 & & & & correction range \\
\hline
single & 1 & $O(\epsilon^0)$ & $\frac12$ & $ [0.99991\pi, 1.00009\pi] $ \\
X3 & 3 & $O(\epsilon)$ & $\frac16,\frac56$ & $ [0.992\pi, 1.008\pi] $ \\
X5 & 5 & $O(\epsilon^2)$ & 0.0672, 0.3854, 1.1364 & $ [0,964\pi, 1.036\pi] $ \\
X7 & 7 & $O(\epsilon^3)$ & 0.2560, 1.6839, 0.5933, 0.8306 & $ [0.925\pi, 1.075\pi] $ \\
X9 & 9 & $O(\epsilon^4)$ & 0.3951, 1.2211, 0.7806, 1.9335, 0.4580 & $ [0.883\pi, 1.117\pi] $ \\
X11 & 11 & $O(\epsilon^5)$ & 0.2984, 1.8782, 1.1547, 0.0982, 0.6883, 0.8301 & $ [0.843\pi, 1.157\pi] $ \\
X13 & 13 & $O(\epsilon^6)$ & 0.8800, 0.6048, 1.4357, 0.9817, 0.0781, 0.5025, 1.8904 & $ [0.807\pi, 1.193\pi] $ \\
X15 & 15 & $O(\epsilon^7)$ & 0.5672, 1.4322, 0.9040, 0.2397, 0.9118, 0.5426, 1.6518, 0.1406 & $ [0.773\pi, 1.227\pi] $  \\
X17 & 17 & $O(\epsilon^8)$ & 0.3604, 1.1000, 0.7753, 1.6298, 1.2338, 0.2969, 0.6148, 1.9298, 0.4443 & $ [0.743\pi, 1.257\pi] $  \\
% \normalsize
\hline
\end{tabular}
\caption{
Phases of symmetric composite sequences of $N=2n+1$ nominal $\pi$ pulses, which produce the X gate with a pulse area error compensation up to order $O(\epsilon^n)$.
The last column gives the high-fidelity range $[\pi (1-\epsilon_0), \pi (1+\epsilon_0)]$ of pulse area error compensation wherein the Frobenius distance fidelity is above the value $0.9999$, i.e. the fidelity error is below $10^{-4}$. \\
}
\label{Table:X}
\end{table*}
%T%T%T%T%T%T%T%T%T%T%T%T%T%T%T%T%T%T%T%T%T%T%T%T%T%T%T%T%T%T%T%T%T%T%T%T%T%T%T%T%T%T%T%T%T

The same second-order error compensation, and the same fidelity, can be obtained by composite sequences of five pulses of area $\pi$ each,
\be\label{X-5}
 \pi_{\phi_1} \pi_{\phi_2} \pi_{\phi_3} \pi_{\phi_4} \pi_{\phi_5}.
\ee
Hence the total pulse area is $5\pi$, the same as the four-pulse sequences above.
Because of the additional phase compared to the four-pulse sequences, various phase choices are possible.
For example, an asymmetric sequence of the kind \eqref{X-5} has the phases
$\phi_1=0$,
$\phi_2 = \arcsin \left( \frac{14+\sqrt{31}}{20}\right) \approx 0.4337\pi$,
$\phi_3 = \pi + \arcsin\left(  \frac{9\sqrt{31}-19}{80}\right) \approx 1.1271\pi$,
$\phi_4 = \arcsin\left(  \frac{9\sqrt{31}+19}{80}\right) \approx 0.3320\pi$,
$\phi_5 = \arcsin\left(  \frac{14-\sqrt{31}}{20}\right) \approx 0.1385\pi$.

We have derived also the symmetric sequence
\be\label{X5-sym}
 \pi_{\phi_1} \pi_{\phi_2} \pi_{\phi_3} \pi_{\phi_2} \pi_{\phi_1},
\ee
with $\phi_1 = \arcsin\left(1-\sqrt{5/8}\right) \approx 0.0672\pi$, $\phi_2 = \arcsin\left((3\sqrt{10}-2)/8\right) \approx 0.3854\pi$, $\phi_3 = 2\phi_2 - 2\phi_1 + \pi/2 \approx 1.1364\pi$.
For these five-pulse sequences the Frobenius infidelity $\mathcal{I}_2$ is given again by Eq.~\eqref{F-X-4}, and the trace infidelity by $\mathcal{I}_2^2$.
The respective fidelities are plotted in Fig.~\ref{fig:X135}.
Obviously, they are much larger than the respective fidelities for a single pulse and the three-pulse composite sequence \eqref{X3-phases}.

The Frobenius distance infidelity \eqref{F-X-4} remains below the quantum computation fidelity threshold $10^{-4}$ in the pulse area interval $(0.964\pi, 1.036\pi)$, i.e. for relative errors up to $|\epsilon| < 0.036$.
On the other hand, the trace distance infidelity $\mathcal{I}_2^2$ remains above this value in the pulse area interval $(0.832\pi, 1.168\pi)$, i.e. for relative errors up to $|\epsilon| < 0.168$, a factor of almost 5 larger.
As for the three-pulse composite X gate, as seen in Fig.~\ref{fig:X135}, the Frobenius distance fidelity is much more demanding error measure as its error is much larger than the error of the trace distance fidelity.

Hereafter we will leave out the trace distance fidelity \eqref{trace fidelity} and will use only the Frobenius distance fidelity \eqref{Frobenius}, because it is a much stricter measure of the gate error.

We conclude this subsection by noting that the availability of various four- and five-pulse symmetric and asymmetric sequences which produce the same fidelity is not a redundancy because they may have rather different sensitivity to phase errors, as has been shown recently for other composite sequences \cite{Torosov2019phases}.

\subsection{Higher-order error compensation}\label{Sec:X-more}

For composite sequences of more than 5 pulses, the equations for the composite phases quickly become very cumbersome and impossible to solve analytically.
They repeat the pattern of the sequences of four and five pulses above: the composite sequences of $2n$ and $2n+1$ pulses have a total pulse area of $(2n+1)\pi$, with all pulses in the sequence being nominal $\pi$ pulses, with the exception of one of the pulses in the $2n$-pulse sequence which has a nominal pulse area of $2\pi$.
Either sequences of $2n$ and $2n+1$ pulses produce error compensation of the order $O(\epsilon^n)$ and identical fidelity profiles.

%***************************************************************
\begin{figure}[t]
\bt{r}
\includegraphics[width=0.947\columnwidth]{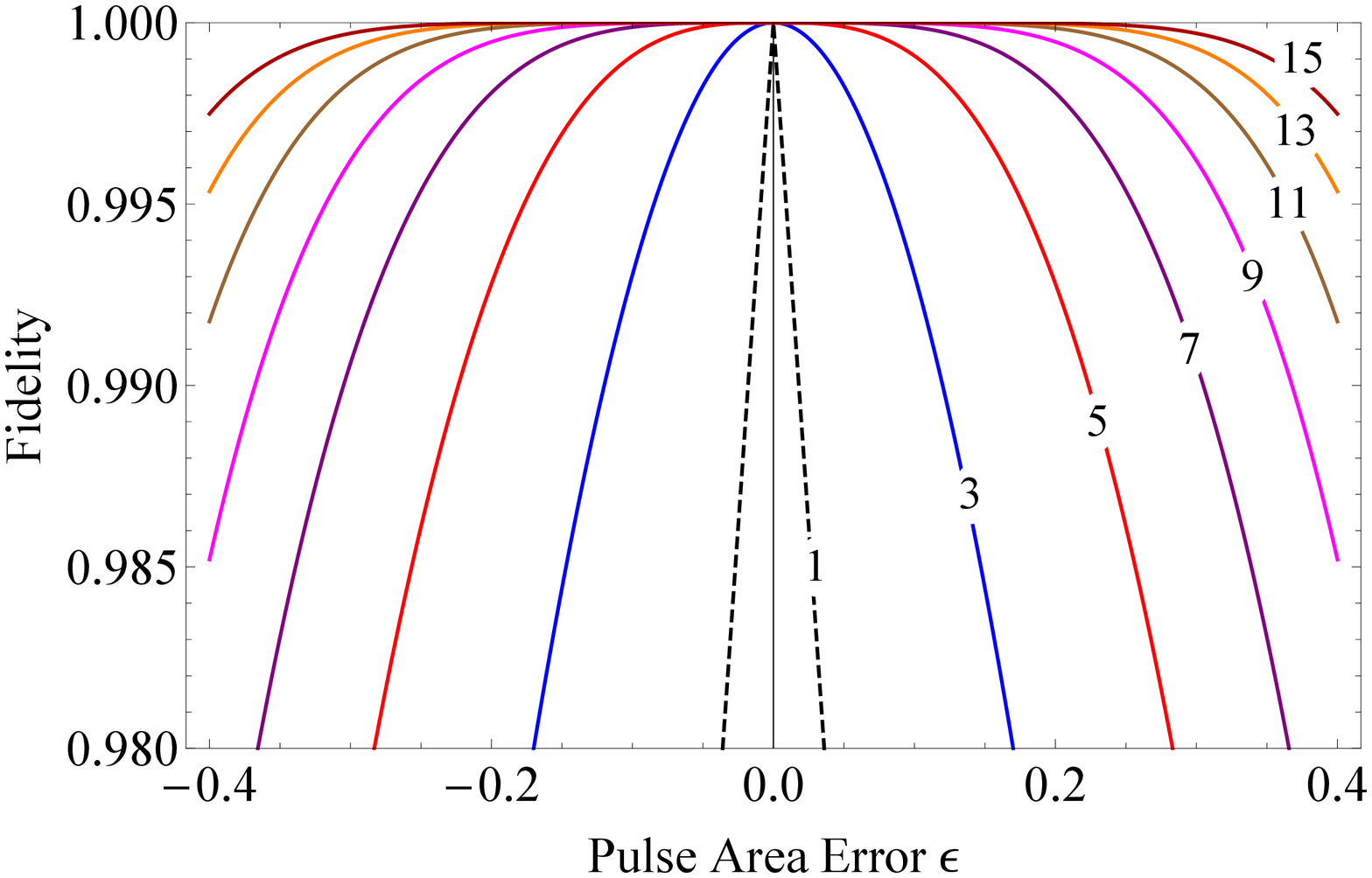} \\ \\
\includegraphics[width=0.93\columnwidth]{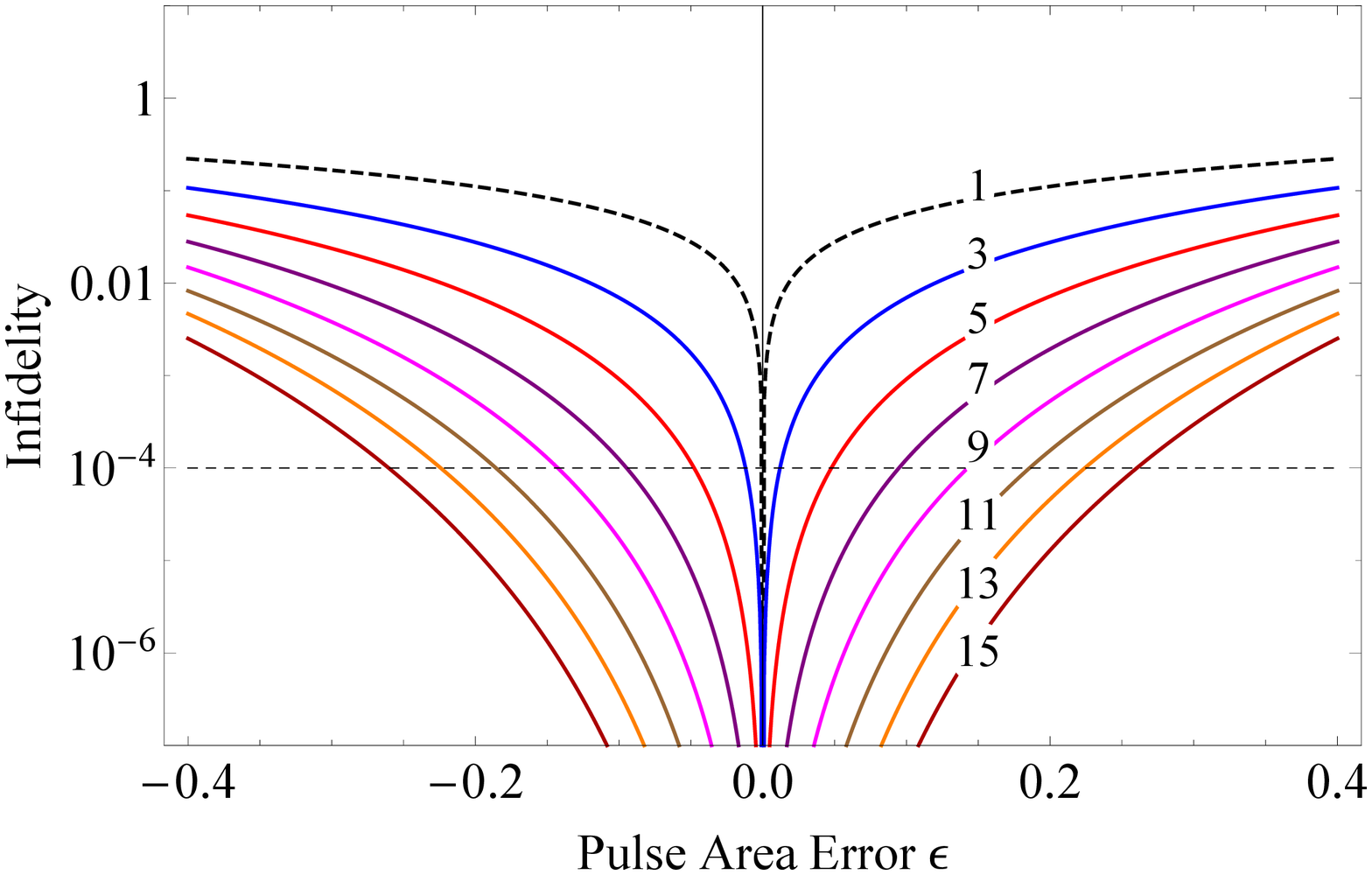}
\et
\caption{
Frobenius distance fidelity $F$ (top) and infidelity (bottom) of composite X gates.
The infidelity is in logarithmic scale in order to better visualize the high-fidelity (low-infidelity) range.
The numbers $N$ on the curves refer to composite sequences X$N$ listed in Table \ref{Table:X}.
}
\label{fig:X}
\end{figure}
%***************************************************************

The $2n+1$-pulse sequences have an additional free phase which can be used to make the composite sequence \emph{symmetric} as in Eq.~\eqref{CP-symmetric}, viz.
\be\label{X-symmetric}
\pi_{\phi_1} \pi_{\phi_2} \pi_{\phi_3} \cdots \pi_{\phi_{n-1}} \pi_{\phi_n} \pi_{\phi_{n-1}} \cdots \pi_{\phi_3} \pi_{\phi_2} \pi_{\phi_1}.
\ee
The propagators generated by the symmetric composite sequences \eqref{X-symmetric} feature two important properties:
\begin{enumerate}
\item All even-order derivatives $\mathcal{U}^{(2k)}_{11}(0)$ of the diagonal elements in Eq.~\eqref{eq-m} vanish, and so do all odd-order derivatives $\mathcal{U}^{(2k+1)}_{12}(0)$ of the off-diagonal elements.
\item The remaining nonzero derivatives in Eq.~\eqref{eq-m} are either real or imaginary: $\mathcal{U}^{(2k+1)}_{11}(0)$ are real, whereas $\mathcal{U}^{(2k)}_{12}(0)$ are imaginary.
\end{enumerate}
Therefore, Eqs.~\eqref{eq-0} and \eqref{eq-m} reduce to a set of $n+1$ real trigonometric equations for $n+1$ free phases.
There are multiple solutions for the phases for every $(2n+1)$-pulse composite sequence.

Two of the phases can be found analytically.
The solution of the zeroth-order Eqs.~\eqref{eq-0} reads
\be
\phi_{n+1} = \frac{\pi}{2} +2 [\phi_{n} - \phi_{n-1} + \phi_{n-2} - \phi_{n-3} +\cdots + (-)^{n} \phi_1].
\ee
Given this relation, the equation $\mathcal{U}^{(1)}_{11}(0) = 0$ reduces to
\be
2 \sum_{k=1}^{n} \sin (\Phi_k) = (-)^{n+1},
\ee
with
\begin{align}
\Phi_k &= 2\sum_{j=1}^{k-1} (-)^{j+1} \phi_j + (-)^{k+1} \phi_k \notag\\
&= 2[\phi_1 - \phi_2 + \phi_3 + \cdots + (-)^{k} \phi_{k-1}] + (-)^{k+1} \phi_k,
\end{align}
from where we can find $\phi_n$.
For example, for 3, 5, and 7 pulses we have, respectively,
\bse
\begin{align}
&\sin (\phi_1) + \sin (2\phi_1 - \phi_2) = -\tfrac12, \\
&\sin (\phi_1) + \sin (2\phi_1 - \phi_2) + \sin (2\phi_1 - 2\phi_2 + \phi_3) = \tfrac12, \\
&\sin (\phi_1) + \sin (2\phi_1 - \phi_2) + \sin (2\phi_1 - 2\phi_2 + \phi_3) \notag\\
&\qquad + \sin (2\phi_1 - 2\phi_2 + 2 \phi_3 - \phi_4) = -\tfrac12.
\end{align}
\ese
From each of these we can find two solutions for the phase with the largest subscript.

The remaining $n-1$ phases $\phi_1, \phi_2, \ldots, \phi_{n-1}$ can be determined numerically.

We have derived numerically the composite phases of symmetric sequences of an odd number of pulses, Eq.~\eqref{X-symmetric}.
They are presented in Table \ref{Table:X}.
The fidelity of these composite X gates is plotted in Fig.~\ref{fig:X}.
It is clear from the table and the figure that a single pulse has very little room for errors as the high-fidelity X gate allows for pulses area errors of less than 0.01\%.
The three-pulse composite X gate offers some leeway, with the admissible error of 0.8\%.
The real pulse area error correction effect is achieved with the composite sequences of 5 to 9 pulses, for which the high-fidelity range of admissible errors increases from 3.6\% to 11.7\%.
Quite remarkably, errors of up to 25\% can be eliminated, and ultrahigh fidelity maintained, with the 17-pulse composite X gate.
Note that these error ranges are calculated by using the rather tough Frobenius distance fidelity \eqref{Frobenius}.
Had we use the much more relaxed trace distance fidelity \eqref{trace fidelity}, these ranges would be much broader, see the numbers for 1, 3 and 5 pulses above.

That said, very long sequences are barely practical because the gate is much slower.
Moreover, it is hard to imagine a quantum computer operating with 25\% pulse area error.
Therefore, the composite sequences of 5, 7 and 9 pulses seems to offer the best fidelity-to-speed ratio.

\section{Hadamard gate}\label{Sec:H}

We shall use the following form of the Hadamard gate (known as pseudo-Hadamard form),
\be\label{H-gate}
\mathbf{H} = \mathbf{R}_y(\pi/2) = e^{i (\pi/4) \hat\sigma_y } = \tfrac{1}{\sqrt{2}} \left[ \begin{array}{cc} 1 & 1 \\  -1 & 1 \end{array} \right].
\ee
It is SU(2) symmetric and it is  equivalent to the more common Walsh-Hadamard form
\be\label{H-gate-1}
 \tfrac{1}{\sqrt{2}} \left[ \begin{array}{cc} 1 & 1 \\  1 & -1 \end{array} \right],
\ee
which is not SU(2) symmetric.
The gate \eqref{H-gate} is equivalent to the often used SU(2) symmetric gate (known as the Splitter gate)
\be\label{H-gate-i}
\mathbf{H}_x = e^{i (\pi/4) \hat\sigma_x } = \tfrac{1}{\sqrt{2}} \left[ \begin{array}{cc} 1 &  i \\   i & 1 \end{array} \right],
\ee
which is related to it by a phase transformation.

The Hadamard gate can be generated by an ideal resonant $\pi/2$ pulse, which is, however, prone to experimental errors.
In order to construct the composite Hadamard gate we have considered all three types of composite sequences \eqref{CP-symmetric}, \eqref{CP-asymmetric-theta}, and \eqref{CP-asymmetric-ab}.
Below we consider these sequences, in the increasing order of error compensation.

%T%T%T%T%T%T%T%T%T%T%T%T%T%T%T%T%T%T%T%T%T%T%T%T%T%T%T%T%T%T%T%T%T%T%T%T%T%T%T%T%T%T%T%T%T
\begin{table*}
\begin{tabular}{|c|c|c|l|l|r|c|%%
}
\hline
 \multicolumn{7}{|c|}{Symmetric sequences $\alpha_{\phi_1} \pi_{\phi_2} \cdots  \pi_{\phi_n} \pi_{\phi_{n+1}} \pi_{\phi_n} \cdots \pi_{\phi_2} \alpha_{\phi_1}$} \\
\hline
notation & N & $O(\epsilon^n)$ & $\alpha$ & $\phi_1,\phi_2, \ldots, \phi_n$ (in units $\pi$) & $\mathcal{A}_{\text{tot}}$ & Range \\
\hline
H3s & 3 & $O(\epsilon)$ & 0.6399 & 1.8442, 1.0587 & $2.28\pi$ & $[0.988,1.012]\pi$ \\
H5s & 5 & $O(\epsilon^2)$ & 0.45 & 1.9494, 0.5106, 1.3179 & $3.90\pi$ & $[0.952,1.048]\pi$ \\
H7s & 7 & $O(\epsilon^3)$ & 0.2769 &  1.6803, 0.2724, 0.8255, 1.6624 & $5.55\pi$ & $[0.905,1.095]\pi$ \\
H9s & 9 & $O(\epsilon^4)$ & 0.2947 & 0.2711, 1.1069, 1.5283, 0.1283, 0.9884 & $7.59\pi$ & $[0.857,1.143]\pi$ \\
H11s & 11 & $O(\epsilon^5)$ & 0.2985 & 1.7377, 0.1651, 0.9147, 0.1510, 0.9331, 1.6415 & $9.60\pi$ & $[0.814,1.186]\pi$ \\
H13s & 13 & $O(\epsilon^6)$ & 0.5065 & 0.0065, 1.7755, 0.7155, 0.5188, 0.2662, 1.2251, 1.3189 & $12.01\pi$ & $[0.776,1.224]\pi$ \\
H15s & 15 & $O(\epsilon^7)$ & 0.3213 & 1.2316, 0.9204, 0.2043, 1.9199, 0.8910, 0.7381, 1.9612, 1.3649 & $13.64\pi$ & $[0.740,1.260]\pi$ \\ \hline
\multicolumn{7}{|c|}{Asymmetric sequences $(\pi/2)_{\phi_1} \pi_{\phi_2} \pi_{\phi_3} \cdots \pi_{\phi_{N-1}} \pi_{\phi_N}$} \\
\hline
notation & N & $O(\epsilon^n)$ & $\alpha$, $\beta$ & $\phi_1,\phi_2, \ldots, \phi_N$ (in units $\pi$) & $\mathcal{A}_{\text{tot}}$ & Range \\
\hline
H5w & 5 & $O(\epsilon^2)$ & 0.5, 1.0 & 0.5, 1.0399, 0.1197, 0.1197, 1.0399 & $4.50\pi$ & $[0.952,1.048]\pi$ \\
H7w & 7 & $O(\epsilon^3)$ & 0.5, 1.0 & 0.5, 1.4581, 0.7153, 0.1495, 1.3738, 0.2568, 0.7752 & $6.50\pi$ & $[0.905,1.095]\pi$ \\
H9w & 9 & $O(\epsilon^4)$ & 0.5, 1.0 & 0.5, 1.1990, 0.3622, 0.6007, 1.6773, 1.7779, 0.6773, 04124, 1.2732 & $8.50\pi$ & $[0.857,1.143]\pi$ \\
H11w & 11 & $O(\epsilon^5)$ & 0.5, 1.0 &0.5, 0.7807, 0.1769, 1.4678, 0.1085, 1.0174, 0.2988, 0.8883, & & \\
 & & & &  \quad 1.2697, 0.3773, 1.6775   & $10.50\pi$ & $[0.814,1.186]\pi$ \\
H13w & 13 & $O(\epsilon^6)$ & 0.5, 1.0 & 0.5, 1.3795, 0.5435, 0.5111, 1.3032, 0.4295, 1.7578, 1.4181, & & \\
 & & & &  \quad  0.3340, 0.4403, 1.7563, 0.6708, 1.1544 & $12.50\pi$ & $[0.776,1.224]\pi$ \\
\hline
 \multicolumn{7}{|c|}{Asymmetric sequences $\alpha_{\phi_1} \pi_{\phi_2} \pi_{\phi_3} \cdots \pi_{\phi_{N-1}} \beta_{\phi_N}$} \\
\hline
notation & N & $O(\epsilon^n)$ & $\alpha$, $\beta$ & $\phi_1,\phi_2, \ldots, \phi_N$ (in units $\pi$) & $\mathcal{A}_{\text{tot}}$ & Range \\
\hline
H4a & 4 & $O(\epsilon^2)$ & 0.7821, 1.3914 & 1.8226, 0.6492, 1.2131, 0.3071 & $4.17\pi$ & $[0.952,1.048]\pi$ \\
H6a & 6 & $O(\epsilon^3)$ & 0.5917, 1.1305 & 1.5943, 0.2860, 0.8435, 1.6553, 0.7962, 0.2523 & $5.72\pi$ & $[0.905,1.095]\pi$ \\
H8a & 8 & $O(\epsilon^4)$ & 0.4954, 0.9028 & 1.5971, 0.7674, 0.5721, 1.8487, 1.0592, 1.9512, 0.3824, 0.9846 & $7.40\pi$  & $[0.857,1.143]\pi$ \\
H10a & 10 & $O(\epsilon^5)$ & 0.6041, 1.1819 & 1.3480, 0.9259, 0.0292, 0.7288, 0.0996, 1.3909, 0.0183, 0.9322, & & \\
 & & & &  \quad 0.2169, 0.7975  & $9.79\pi$ & $[0.814,1.186]\pi$ \\
H12a & 12 & $O(\epsilon^6)$ & 0.4168, 0.8841 & 1.5817, 1.1160, 0.3751, 0.9583, 0.1333, 1.9445, 1.0381, 1.6293, & & \\
  & & & &  \quad 0.4845, 0.0046, 0.8278, 0.7416  & $11.30\pi$ & $[0.776,1.224]\pi$ \\
\hline
\end{tabular}
\caption{
Phases of three types of composite sequences, which produce the Hadamard gate with a pulse area error compensation up to order $O(\epsilon^n)$.
The total pulse area $\mathcal{A}_{\text{tot}}$ and the high-fidelity range $[\pi-\epsilon_0,\pi+\epsilon_0]$ wherein the Frobenius distance infidelity remains below $10^{-4}$ are listed in the last two columns.
}
\label{Table:Hadamard}
\end{table*}
%T%T%T%T%T%T%T%T%T%T%T%T%T%T%T%T%T%T%T%T%T%T%T%T%T%T%T%T%T%T%T%T%T%T%T%T%T%T%T%T%T%T%T%T%T

\subsection{First-order error correction}

The shortest pulse sequence that can provide a first-order error compensated Hadamard gate consists of three pulses,
\be\label{H3}
\alpha_{\phi_1} \pi_{\phi_2} \alpha_{\phi_1}.
\ee
Equations \eqref{eq-0} result in the equations
\bse
\begin{align}
&-\sin (\alpha) \cos (\phi _1-\phi _2) = \tfrac 1{\sqrt{2}}, \label{H3-1} \\
&e^{-i \phi _1} \left[ \sin (\phi _1-\phi _2) - i \cos (\alpha) \cos (\phi _1-\phi _2) \right] = \tfrac 1{\sqrt{2}}. \label{H3-2}
\end{align}
The first-derivatives of Eqs.~\eqref{eq-m} are annulled by the single equation
\be
2 \alpha \cos (\phi _1-\phi _2) + 1 = 0. \label{H3-3}
\ee
\ese
From Eqs.~\eqref{H3-1} and \eqref{H3-3} we find
\be
\frac{\sin \alpha}{\alpha} =\sqrt{2}. \label{H3-A}
\ee
Therefore the value of the pulse area $\alpha$ is given by an inverse sinc function of $\sqrt{2}$, which gives $\alpha\approx 0.6399\pi$.
Given $\alpha$, we can find $\phi _1-\phi _2$ from Eq.~\eqref{H3-1} or \eqref{H3-3}, and then $\phi_1$ from
\be
\sqrt{2}\, \sin (\phi _1-\phi _2) = \cos (\phi _1),
\ee
which is the real part of Eq.~\eqref{H3-2}.
The values are $\phi_1 \approx 1.8442\pi$ and $\phi_2 \approx 1.0587\pi$.
Therefore, this composite pulse reads
\be
(0.6399\pi)_{1.8442\pi} \pi_{1.0587\pi} (0.6399\pi)_{1.8442\pi}.
\ee
In term of degrees, it reads ${115^\circ}_{332^\circ} {180^\circ}_{191^\circ} {115^\circ}_{332^\circ}$.
This composite sequence is related to the well-known sequence SCROFULOUS \cite{Cummins2003}.

\subsection{Second-order error correction}

Second-order error compensation is obtained by a composite sequence of at least 4 pulses.
A popular CP is the BB1 pulse of Wimperis \cite{Wimperis1994},
\be\label{BB1-4}
\text{BB1} = (\pi/2)_{0} \pi_{\chi} (2\pi)_{3\chi} \pi_{\chi},
\ee
which produces the gate \eqref{H-gate-i}, with a total pulse area of $4.5\pi$.
It can be viewed as identical to the five-pulse sequence
\be\label{BB1-5}
%\text{H5w} =
(\pi/2)_{0} \pi_{\chi} \pi_{3\chi} \pi_{3\chi} \pi_{\chi}.
\ee

We have derived a different, asymmetric four-pulse CP,
\be\label{H4a}
\text{H4a} = \alpha_{\phi_1} \pi_{\phi_2} \pi_{\phi_3} \beta_{\phi_4},
\ee
where $\alpha = 0.7821\pi$, $\beta = 1.3914\pi$, $\phi_1 = 1.8226\pi$, $\phi_2 = 0.6492\pi$, $\phi_3 = 1.2131\pi$, $\phi_4=0.3071\pi$.
This pulse has a total area of about $4.17\pi$, i.e. it is faster than the BB1 pulse.
It is accurate up to the same order $O(\epsilon^2)$ and produces essentially the same fidelity profile as BB1.

We have also derived a five-pulse composite Hadamard gate by using the symmetric sequence
\be\label{H5s}
\text{H5s} = \alpha_{\phi_1} \pi_{\phi_2} \pi_{\phi_3} \pi_{\phi_2} \alpha_{\phi_1},
\ee
with $\alpha = 0.45\pi$, $\phi_1 = 1.9494\pi$,  $\phi_2 = 0.5106\pi$,  $\phi_3 = 1.3179\pi$.
It delivers again the second-order error compensation $O(\epsilon^2)$, however, with a total pulse area of just about $3.9\pi$.
Therefore it is considerably faster than the BB1 pulse, by over 13\%, while having a similar performance.

%***************************************************************
\begin{figure}[t]
\bt{r}
\includegraphics[width=0.947\columnwidth]{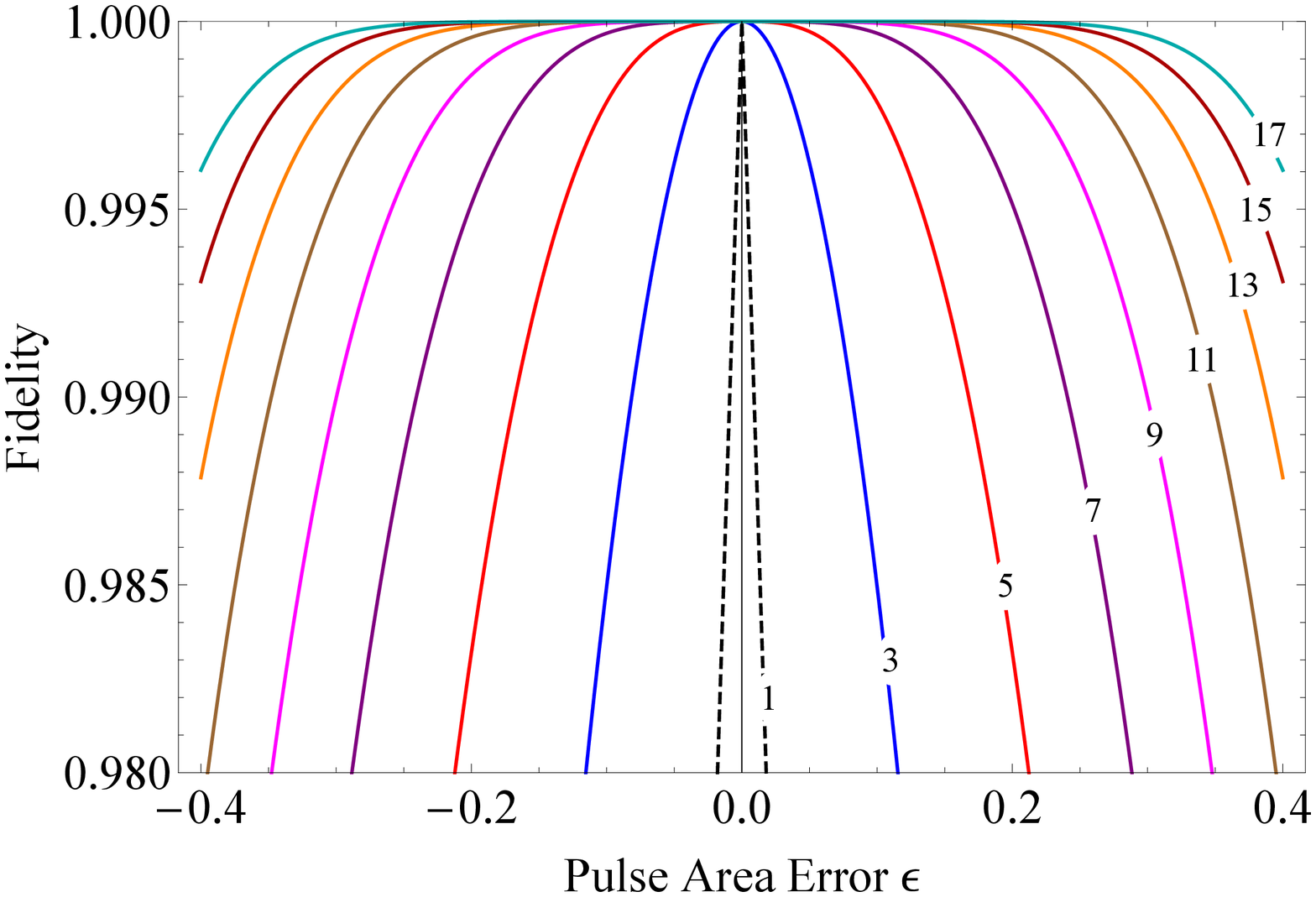} \\ \\
\includegraphics[width=0.93\columnwidth]{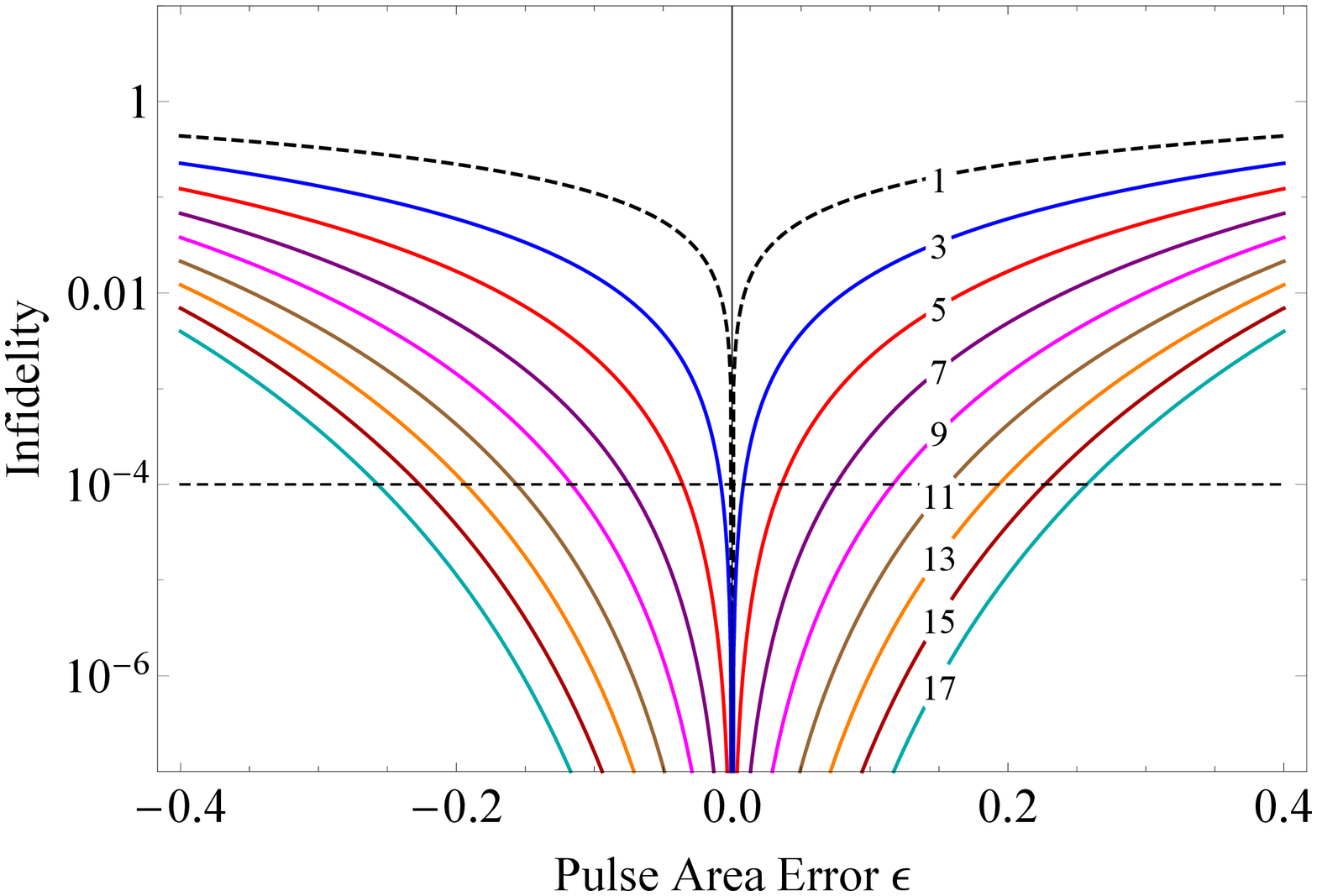}
\et
\caption{
Frobenius distance fidelity (top) and infidelity (bottom) of composite Hadamard gates produced by using the symmetric composite sequences H$N$s from Table \ref{Table:Hadamard}.
}
\label{fig:Hadamard}
\end{figure}
%***************************************************************

%========================

\subsection{Higher-order error correction}

Similarly to the second order, the \emph{third-order error compensation} is obtained in several different manners, requiring at least 6 pulses.
The 6-pulse sequence with the minimal pulse area of about $5.72\pi$ reads
\be\label{H6a}
\text{H6a} = \alpha_{\phi_1} \pi_{\phi_2} \pi_{\phi_3} \pi_{\phi_4} \pi_{\phi_5} \beta_{\phi_6},
\ee
with
$\alpha=0.5917\pi$, $\beta=1.1305\pi$, and the phases given in Table \ref{Table:Hadamard}.
The same error correction order is achieved with the symmetric seven-pulse sequence
\be\label{H7s}
\text{H7s} = \alpha_{\phi_1} \pi_{\phi_2} \pi_{\phi_3} \pi_{\phi_4} \pi_{\phi_3} \pi_{\phi_2} \alpha_{\phi_1},
\ee
with
$\alpha = 0.2769\pi$, and the phases given in Table \ref{Table:Hadamard}.
It produces the same fidelity profile as the 6-pulse sequence but it is a little faster as its pulse area is about $5.55\pi$.
Another seven-pulse composite sequence is built similarly to the BB1 sequence \eqref{BB1-4},
\be\label{BB1-7}
\text{H7w} = (\pi/2)_{\pi/2} \pi_{\phi_2} \pi_{\phi_3} \pi_{\phi_4} \pi_{\phi_5} \pi_{\phi_6} \pi_{\phi_7},
\ee
with the phases given in Table \ref{Table:Hadamard}.
It achieves the same error order compensation $O(\epsilon^3)$, however, with a larger total pulse area of $6.5\pi$ compared to the previous two CPs.

%\subsection{Fourth-order error correction}

%=================

\emph{Fourth-order error compensation} is obtained by at least 8 pulses.
The 8-pulse sequence with the minimal pulse area of about $7.40\pi$ reads
\be
\text{H8a} = \alpha_{\phi_1} \pi_{\phi_2} \pi_{\phi_3} \pi_{\phi_4} \pi_{\phi_5} \pi_{\phi_6} \pi_{\phi_7} \beta_{\phi_8},
\ee
with
$\alpha=0.4954\pi$, $\beta=0.9028\pi$, and the phases are given in Table \ref{Table:Hadamard}.
The same error correction order is achieved with the symmetric nine-pulse sequence
\be
\text{H9s} = \alpha_{\phi_1} \pi_{\phi_2} \pi_{\phi_3} \pi_{\phi_4} \pi_{\phi_5} \pi_{\phi_4} \pi_{\phi_3} \pi_{\phi_2} \alpha_{\phi_1},
\ee
with
$\alpha = 0.2947$, with the phases in Table \ref{Table:Hadamard}.
Its total pulse area is $7.59\pi$.
The BB1-like nine-pulse composite sequence,
\be\label{BB1-9}
\text{H9w} = (\pi/2)_{\pi/2} \pi_{\phi_2} \pi_{\phi_3} \pi_{\phi_4} \pi_{\phi_5} \pi_{\phi_6} \pi_{\phi_7} \pi_{\phi_8} \pi_{\phi_9},
\ee
with the phases in Table \ref{Table:Hadamard}, achieves the same fourth-order error compensation $O(\epsilon^4)$, however, with the largest total pulse area of $8.5\pi$ compared to the previous two CPs.

%=================

The same pattern is repeated for the longer pulse sequences presented in Table \ref{Table:Hadamard}: for the same order of pulse area error compensation, the fastest sequences, with the smallest total pulse area are either the asymmetric H$N$a or symmetric H$N$s sequences, and the BB1-like sequences H$N$w are the slowest ones.

The fidelity and the infidelity of the composite Hadamard gates of up to seventh-order error compensation are plotted in Fig.~\ref{fig:Hadamard}.
Obviously, as the number of pulses in the composite sequences, and hence the compensated error order, increase the fidelity and infidelity profiles improve and get broader.

\section{General rotation gate}\label{Sec:rotation}

\subsection{First-order error correction \label{Sec:rot-symmetric}}

%T%T%T%T%T%T%T%T%T%T%T%T%T%T%T%T%T%T%T%T%T%T%T%T%T%T%T%T%T%T%T%T%T%T%T%T%T%T%T%T%T%T%T%T%T
\begin{table*}
\begin{tabular}{|c|c|c|c|c|}
\hline
{} & 3 pulses, $O(\epsilon)$ & 5 pulses, $O(\epsilon^2)$  & 7 pulses, $O(\epsilon^3)$ & 9 pulses, $O(\epsilon^4)$ \\
 & $\alpha_{\phi_1} \pi_{\phi_2} \alpha_{\phi_1}$ & $\alpha_{\phi_1} \pi_{\phi_2} \pi_{\phi_3} \pi_{\phi_2} \alpha_{\phi_1}$ &
 $\alpha_{\phi_1} \pi_{\phi_2} \pi_{\phi_3} \pi_{\phi_4} \pi_{\phi_3} \pi_{\phi_2} \alpha_{\phi_1}$ &
  $\alpha_{\phi_1} \pi_{\phi_2} \pi_{\phi_3} \pi_{\phi_4} \pi_{\phi_5} \pi_{\phi_4} \pi_{\phi_3} \pi_{\phi_2} \alpha_{\phi_1}$ \\ \hline
$\theta$ &
 $\alpha;\phi_1,\phi_2$ &
 $\alpha;\phi_1,\phi_2, \phi_3$ &
 $\alpha;\phi_1,\phi_2, \phi_3, \phi_4$ &
 $\alpha;\phi_1,\phi_2, \phi_3, \phi_4, \phi_5$ \\
\hline
$\frac1{10}$ & \scriptsize 0.5061; 1.0389, 1.9892 & \scriptsize 0.4548; 0.6416, 1.5230, 0.4258 & \scriptsize 0.4625; 0.7317, 1.8366, 1.0783, 0.1821 & \scriptsize 0.5125; 1.9200, 0.8412, 1.5473, 0.2812, 1.1816 \\
$\frac18$ & \scriptsize 0.5096; 1.0483, 1.9865 & \scriptsize 0.4453; 0.6626, 1.5245, 0.4168 & \scriptsize 0.4500; 0.7069, 1.8222, 1.0860, 0.1970 & \scriptsize 0.5101; 1.9490, 0.8687, 1.5489, 0.2665, 1.1618 \\
$\frac16$ & \scriptsize 0.5169; 1.0636, 1.9819 & \scriptsize 0.4315; 0.6964, 1.5259, 0.4032 & \scriptsize 0.4277; 0.6691, 1.8020, 1.0976, 0.2183 & \scriptsize 0.5022; 1.9918, 0.9092, 1.5502, 0.2455, 1.1340 \\
$\frac15$ & \scriptsize 0.5242; 1.0754, 1.9782 & \scriptsize 0.4225; 0.7231, 1.5263, 0.3934 & \scriptsize 0.4090; 0.6404, 1.7886, 1.1061, 0.2334 & \scriptsize 0.4926; 0.0229, 0.9382, 1.5502, 0.2308, 1.1148 \\
$\frac14$ & \scriptsize 0.5375; 1.0921, 1.9726 & \scriptsize 0.4129; 0.7630, 1.5259, 0.3796 & \scriptsize 0.3803; 0.5977, 1.7717, 1.1181, 0.2536 & \scriptsize 0.4729; 0.0661, 0.9770, 1.5491, 0.2110, 1.0894 \\
$\frac13$ & \scriptsize 0.5653; 1.1173, 1.9628 & \scriptsize 0.4087; 0.8293, 1.5231, 0.3583 & \scriptsize 0.3336; 0.5212, 1.7505, 1.1370, 0.2836 & \scriptsize 0.4269; 0.1326, 1.0314, 1.5448, 0.1815, 1.0525 \\
$\frac12$ & \scriptsize 0.6399; 1.1558, 1.9413 & \scriptsize 0.4500; 0.9494, 1.5106, 0.3179 & \scriptsize 0.2769; 0.3197, 1.7275, 1.1745, 0.3376 & \scriptsize 0.2947; 0.2711, 1.1069, 1.5283, 0.1283, 0.9884 \\
$\frac23$ & \scriptsize 0.7365; 1.1779, 1.9155 & \scriptsize 0.5563; 1.0329, 1.4886, 0.2746 & \scriptsize 0.3410; 0.1020, 1.7252, 1.2168, 0.3923 & \scriptsize 0.1700; 0.5700, 1.1449, 1.5009, 0.0735, 0.9254 \\
$\frac34$ & \scriptsize 0.7925; 1.1827, 1.9000 & \scriptsize 0.6322; 1.0585, 1.4728, 0.2498 & \scriptsize 0.4269; 0.0309, 1.7317, 1.2421, 0.4230 & \scriptsize 0.2045; 0.8134, 1.1515, 1.4816, 0.0423, 0.8905 \\
$\frac45$ & \scriptsize 0.8288; 1.1834, 1.8895 & \scriptsize 0.6857; 1.0688, 1.4613, 0.2332 & \scriptsize 0.4947; 0.0017, 1.7386, 1.2595, 0.4436 & \scriptsize 0.2726; 0.9091, 1.1514, 1.4674, 0.0212, 0.8672 \\
$\frac56$ & \scriptsize 0.8542; 1.1829, 1.8819 & \scriptsize 0.7251; 1.0735, 1.4526, 0.2210 & \scriptsize 0.5474; 1.9872, 1.7446, 1.2725, 0.4586 & \scriptsize 0.3336; 0.9507, 1.1495, 1.4564, 0.0055, 0.8501 \\
$\frac78$ & \scriptsize 0.8874; 1.1812, 1.8717 & \scriptsize 0.7795; 1.0770, 1.4401, 0.2044 & \scriptsize 0.6234; 1.9741, 1.7542, 1.2907, 0.4795 & \scriptsize 0.4275; 0.9853, 1.1446, 1.4404, 1.9837, 0.8264 \\
$\frac9{10}$ & \scriptsize 0.9083; 1.1795, 1.8650 & \scriptsize 0.8154; 1.0777, 1.4316, 0.1934 & \scriptsize 0.6759; 1.96887, 1.7613, 1.3030, 0.4935 & \scriptsize 0.4952; 0.9992, 1.1402, 1.4291, 1.9688, 0.8103 \\
% \normalsize
\hline
\end{tabular}
\caption{
Pulse area $\alpha$ and phases of composite pulse sequences which produce rotation gates of angle $\theta$.
The area $\alpha$ and all phases are given in units $\pi$.
The case of $\theta = \frac12 \pi$ repeats the symmetric Hadamard gates already presented in Sec.~\ref{Sec:H}; they are given here for the sake of comparison and completeness.
}
\label{Table:rotation}
\end{table*}
%T%T%T%T%T%T%T%T%T%T%T%T%T%T%T%T%T%T%T%T%T%T%T%T%T%T%T%T%T%T%T%T%T%T%T%T%T%T%T%T%T%T%T%T%T

The shortest pulse sequence that can provide a first-order error compensation, as for the X and Hadamard gates, consists of three pulses,
\be
\alpha_{\phi_1} \pi_{\phi_2} \alpha_{\phi_1}.
\ee
Equations \eqref{eq-0} result in the equations
\bse
\begin{align}
&-\sin (\alpha) \cos (\phi _1-\phi _2) = \cos(\theta/2), \label{R3-1} \\
&e^{-i \phi _1} \left[ \sin (\phi _1-\phi _2) - i \cos (\alpha) \cos (\phi _1-\phi _2) \right] = \sin(\theta/2). \label{R3-2}
\end{align}
The first-derivatives of Eqs.~\eqref{eq-m} are annulled by the single equation
\be
2 \alpha \cos (\phi _1-\phi _2) + 1 = 0. \label{R3-3}
\ee
\ese
From Eqs.~\eqref{R3-1} and \eqref{R3-3} we find
\be
\frac{\sin (\alpha)}{\alpha} = 2\cos(\theta/2). \label{R3-A}
\ee
Therefore the value of the pulse area $\alpha$ is given by an inverse sinc function of $2\cos(\theta/2)$.
Given $\alpha$, we can find $\phi _1-\phi _2$ from Eq.~\eqref{R3-1} or \eqref{R3-3}, and then $\phi_1$ from
\be
\sin (\phi _1-\phi _2) = \sin (\theta/2) \cos (\phi _1),
\ee
which is obtained from Eq.~\eqref{R3-2}.

This composite sequence is related to the SCROFULOUS composite pulse \cite{Cummins2003}, as mentioned above.
The values of the pulse area and the composite phases are given in Table \ref{Table:rotation}.

\subsection{More than three pulses}

The five-pulse sequence,
\be
\alpha_{\phi_1} \pi_{\phi_2} \pi_{\phi_3} \pi_{\phi_2} \alpha_{\phi_1},
\ee
provides a second-order error compensation.
The sequences with 7, 9, etc. pulses have the same structure and deliver an error compensation of order 3, 4, etc.
Generally, a $2n+1$-pulse symmetric sequence of this structure delivers an error compensation up to order $O(\epsilon^n)$.
Unfortunately, analytic expressions for the composite parameters for more than three pulses are hard to obtain, if possible at all.
Hence we have derived them numerically and their values are listed in Table \ref{Table:rotation}.
The fidelity of these sequences behave similarly to the ones for the X and Hadamard gates.

%%%%%%%%%%%%%%%%%%%%%%%%%%%%%%%%%%%%%%%%%%%%%%%%%%%%%%%%%%%%%%%%%%%%%%%%%%%%%%%%%%%%%%%%%%%%%%%%%%%%%%%%%%%%%%%%%%%%%%%%%%%%%%%%%%%%%%%%%
%%%%%%%%%%%%%%%%%%%%%%%%%%%%%%%%%%%%%%%%%%%%%%%%%%%%%%%%%%%%%%%%%%%%%%%%%%%%%%%%%%%%%%%%%%%%%%%%%%%%%%%%%%%%%%%%%%%%%%%%%%%%%%%%%%%%%%%%%
\section{Comments and conclusions\label{Sec:conclusion}}
%%%%%%%%%%%%%%%%%%%%%%%%%%%%%%%%%%%%%%%%%%%%%%%%%%%%%%%%%%%%%%%%%%%%%%%%%%%%%%%%%%%%%%%%%%%%%%%%%%%%%%%%%%%%%%%%%%%%%%%%%%%%%%%%%%%%%%%%%
%%%%%%%%%%%%%%%%%%%%%%%%%%%%%%%%%%%%%%%%%%%%%%%%%%%%%%%%%%%%%%%%%%%%%%%%%%%%%%%%%%%%%%%%%%%%%%%%%%%%%%%%%%%%%%%%%%%%%%%%%%%%%%%%%%%%%%%%%

In this work we presented a number of composite pulse sequences for three basic quantum gates --- the X gate, the Hadamard gate and arbitrary rotation gates.
The composite sequences contain up to 17 pulses and can compensate up to eight orders of experimental errors in the pulse amplitude and duration.
The short composite sequences are calculated analytically and the longer ones numerically.

Three classes of composite sequences have been derived --- one symmetric and two asymmetric.
For the X gate, the three classes coalesce into a single set of symmetric sequences of nominal $\pi$ pulses presented in Table \ref{Table:X}.
For the Hadamard gate, cf. Table \ref{Table:Hadamard}, two of the classes contain as their lowest members two well-known composite sequences: the three-pulse symmetric SCROFULOUS pulse \cite{Cummins2003} and the four-pulse asymmetric BB1 pulse \cite{Wimperis1994}, which compensate first and second-order pulse area errors, respectively.
The third, asymmetric class of composite sequences, does not contain members published before.
All three classes produce essentially identical fidelity profiles for the same order of error compensation.
In general, the SCROFULOUS-like symmetric sequences H$N$s and the asymmetric sequences H$N$a require the least total pulse area and hence are the fastest, whereas the asymmetric BB1-like sequences H$N$w are the slowest.
For the general rotation gates, the three classes behave similarly, although we have presented only the symmetric sequences in Table \ref{Table:rotation} for the sake of brevity.

The composite rotations derived here outperform the existing composite rotations in terms of either speed, or accuracy, or both.
Although we could not improve the first-order  SCROFULOUS sequence, we have derived second-order composite sequences which are faster (by over 13\%) than the famous BB1 sequence \cite{Wimperis1994}:
 our second-order error compensated Hadamard gate has a total nominal pulse area of about $3.9\pi$, which is substantial improvement over the BB1 pulse, which delivers the same error order with a total pulse area of $4.5\pi$ \cite{Wimperis1994}.
The longer composite sequences are derived by brute numerics and they are much shorter than previous sequences with the same order of error compensation obtained by nesting and concatenation of short sequences.
For example, our $n$th order error-compensated X gates are constructed by $2n+1$ nominal $\pi$ pulses, which is much shorter than the concatenated composite sequences.
For example, the 5th order error compensation is produced by a concatenated 15-pulse sequence, whereas we achieve this by an 11-pulse sequence.
Similar scaling applies to the Hadamard and the rotation gates.

The results presented in this work demonstrate the remarkable flexibility of composite pulses accompanied by extreme accuracy and robustness to errors --- three features that cannot be achieved together by any other coherent control technique.
We expect these composite sequences, in particular the X and Hadamard gates, to be very useful quantum control tools in quantum information applications because they provide a variety of options to find the optimal balance between ultrahigh fidelity, error range and speed, which may be different in different physical systems.

%%%%%%%%%%%%%%%%%%%%%%%%%%%%%%%%%%%%%%%%%%%%%%%%%%%%%%%%%%%%%%%%%%%%%%%%%%%%%%%%%%%%%%%%%%%%%%%%%%%%%%%%%%%%%%%%%%%%%%%%%%%%%%%%%%%%%%%%%%%%%%%%%%%%%%%%%%%%%%%%

\acknowledgments
HG acknowledges support from the EU Horizon-2020 ITN project LIMQUET (Contract No. 765075).
NVV acknowledges support from the Bulgarian Science Fund Grant DO02/3 (Quant-ERA Project ERyQSenS).

%%%%%%%%%%%%%%%%%%%%%%%%%%%%%%%%%%%%%%%%%%%%%%%%%%%%%%%%%%%%%%%%%%%%%%%%%%%%%%%%%%%%%%%%%%%%%%%%%%%%%%%%%%%%%%%%%%%%%%%%%%%%%%%%%%%%%%%%%%%%%%%%%%%%%%%%%%%%%%%%
%%%%%%%%%%%%%%%%%%%%%%%%%%%%%%%%%%%%%%%%%%%%%%%%%%%%%%%%%%%%%%%%%%%%%%%%%%%%%%%%%%%%%%%%%%%%%%%%%%%%%%%%%%%%%%%%%%%%%%%%%%%%%%%%%%%%%%%%%%%%%%%%%%%%%%%%%%%%%%%%
%%%%%%%%%%%%%%%%%%%%%%%%%%%%%%%%%%%%%%%%%%%%%%%%%%%%%%%%%%%%%%%%%%%%%%%%%%%%%%%%%%%%%%%%%%%%%%%%%%%%%%%%%%%%%%%%%%%%%%%%%%%%%%%%%%%%%%%%%%%%%%%%%%%%%%%%%%%%%%%%


\begin{thebibliography}{99}

\bibitem{Nielsen2000}
M. A. Nielsen and I. L. Chuang, \textit{Quantum Computation and Quantum Information} (Cambridge University Press, 2000).

\bibitem{Vandersypen2004} L. M. K. Vandersypen and I. L. Chuang, Rev. Mod. Phys. \textbf{76}, 1037 (2004).
%NMR techniques for quantum control and computation

\bibitem{Jones2011}
J. A. Jones, Prog. Nucl. Magn. Res. Spectr. \textbf{59}, 91 (2011).
%Quantum computing with NMR


\bibitem{Vitanov2001} N. V. Vitanov, T. Halfmann, B. W. Shore and K. Bergmann,
	Annu. Rev. Phys. Chem. \textbf{52}, 763 (2001).

\bibitem{Landau1932} L. D. Landau, Phys. Z. Sowjetunion {\bf 2}, 46 (1932).

\bibitem{Zener1932} C. Zener, Proc. R. Soc. Lond. Ser. A {\bf 137}, 696 (1932).

\bibitem{Stuckelberg1932} E. C. G. Stueckelberg, Helv. Phys. Acta {\bf 5}, 369 (1932).

\bibitem{Majorana1932} E. Majorana, Nuovo Cimento {\bf 9}, 43 (1932).


\bibitem{Yatsenko2002} L. P. Yatsenko, N. V. Vitanov, B. W. Shore, T. Rickes, and K. Bergmann, Opt. Commun. \textbf{204}, 413 (2002).

\bibitem{Vitanov2006} N. V. Vitanov and B. W. Shore, Phys. Rev. A \textbf{73}, 053402 (2006).

\bibitem{Yamazaki2008} R. Yamazaki, K.-I. Kanda, F. Inoue, K. Toyoda, and S. Urabe, Phys. Rev. A 78, 023808 (2008).

\bibitem{Zlatanov2017} K. N. Zlatanov and N. V. Vitanov, Phys. Rev. A \textbf{96}, 013415 (2017).

\bibitem{Randall2018} J. Randall, A. M. Lawrence, S. C. Webster, S. Weidt, N. V. Vitanov, and W. K. Hensinger, Phys. Rev. A \textbf{98}, 043414 (2018).

\bibitem{Vitanov2017} N. V. Vitanov, A. A. Rangelov, B. W. Shore, and K. Bergmann, Rev. Mod. Phys. \textbf{89}, 015006 (2017).

\bibitem{Zlatanov2020} K. N. Zlatanov and N. V. Vitanov, Phys. Rev. A \textbf{101}, 013426 (2020).
%Generation of arbitrary qubit states by adiabatic evolution split by a phase jump

\bibitem{Marte1991} P. Marte, P. Zoller, and J. L. Hall, Phys. Rev. A \textbf{44}, R4118 (1991).

\bibitem{Weitz1994} M. Weitz, B. C. Young, and S. Chu, Phys. Rev. Lett. \textbf{73}, 2563 (1994).

\bibitem{Vitanov1999} N. V. Vitanov, K. A. Suominen, and B. W. Shore, J. Phys. B  \textbf{32}, 4535 (1999).

\bibitem{Unanyan1998} R. Unanyan, M. Fleischhauer, B. W. Shore, and K. Bergmann, Opt. Commun. \textbf{155}, 144 (1998).

\bibitem{Theuer1999} H. Theuer, R. G. Unanyan, C. Habscheid, K. Klein, and K. Bergmann, Opt. Express  \textbf{4}, 77 (1999).

\bibitem{Vewinger2003} F. Vewinger, M. Heinz, R. Garcia Fernandez, N. V. Vitanov, and K. Bergmann, Phys. Rev. Lett. \textbf{91}, 213001 (2003).

\bibitem{Ivanov2007} P. A. Ivanov, B. T. Torosov, and N. V. Vitanov, Phys. Rev. A \textbf{75}, 012323 (2007).
%Navigation between quantum states by quantum mirrors

\bibitem{Rousseaux2013}
B. Rousseaux, S. Gu\'erin, N. V. Vitanov, Phys. Rev. A \textbf{87}, 032328	(2013).
%Arbitrary qudit gates by adiabatic passage




	
	\bibitem{Levitt1979}
	M. H. Levitt and R. Freeman, J. Magn. Reson. \textbf{33}, 473 (1979); %NMR population inversion using a composite pulse
	%R. Freeman, \emph{Spin Choreography} (Spektrum, Oxford, 1997).
	
	\bibitem{Freeman1980}
	R. Freeman, S. P. Kempsell, and M. H. Levitt, J. Magn. Reson. \textbf{38}, 453 (1980).
	%Radiofrequency pulse sequences which compensate their own imperfections
	
	\bibitem{Levitt1982}
	M. H. Levitt, J. Magn. Reson. \textbf{48}, 234 (1982).
	%Symmetrical composite pulse sequences for NMR population inversion. I. Compensation of radiofrequency field inhomogeneity
	
	\bibitem{Levitt1983} M. H. Levitt and R. R. Ernst, J. Magn. Res. \textbf{55}, 247 (1983).

	\bibitem{Levitt1986} M. H. Levitt, Prog. NMR Spectrosc. \textbf{18}, 61 (1986).

% \bibitem{Levitt2001} M. H. Levitt, \emph{Spin dynamics: basics of nuclear magnetic resonance} (Wiley, Chichester, 2001).

\bibitem{Levitt2007}
M. H. Levitt, \emph{Composite pulses}, in \emph{Encyclopedia of Magnetic Resonance}, p. 1396 (Wiley, 2007).	


\bibitem{Merrill2014review} J. T. Merrill and K. R. Brown, Adv. Chem. Phys. \textbf{154}, 241 (2014)
%: \emph{Quantum Information and Computation for Chemistry}, edited by Sabre Kais (John Wiley \& Sons, 2014).
% - arXiv preprint arXiv:1203.6392, 2012
%Progress in compensating pulse sequences for quantum computation


\bibitem{Wimperis1990} S. Wimperis, J. Magn. Reson. \textbf{86}, 46 (1990).


\bibitem{Wimperis1991} S. Wimperis, J. Magn. Reson. \textbf{93}, 199 (1991).
%Iterative schemes for phase-distortionless composite 180 pulses,

\bibitem{Wimperis1994} S. Wimperis, J. Magn. Reson. \textbf{109}, 221 (1994).

\bibitem{Cho1986} H.M. Cho, R. Tycko, A. Pines, and J. Guckenheimer, Phys. Rev. Lett. \textbf{56}, 1905 (1986).
%Iterative maps for bistable excitation of two-level systems,

\bibitem{Cho1987} H. Cho, J. Baum, and A. Pines, J. Chem. Phys. \textbf{86}, 3089 (1987).
%Iterative maps with multiple fixed points for excitation of two-level systems,
%[22] R. Tycko, Iterative methods in the design of pulse sequences for NMR excitation, Adv. Magn. Reson. 15 (1990) 1–49.

	
	\bibitem{Torosov2011PRA} B.~T.~Torosov and N. V.~Vitanov, Phys. Rev. A \textbf{83}, 053420 (2011).
	
	\bibitem{Torosov2011PRL} B.~T.~Torosov, S. Gu\'erin and N.~V.~Vitanov, Phys. Rev. Lett. \textbf{106}, 233001 (2011).

\bibitem{Schraft2013} D. Schraft, T. Halfmann, G. T. Genov, and N. V. Vitanov, Phys. Rev. A \textbf{88}, 063406 (2013).
	%Experimental demonstration of composite adiabatic passage

\bibitem{Genov2014} G. T. Genov, D. Schraft, T. Halfmann and N. V. Vitanov, Phys. Rev. Lett. \textbf{113}, 043001 (2014).


	
	\bibitem{Tycko1984} R. Tycko and A. Pines, Chem. Phys. Lett. \textbf{111}, 462 (1984).
	
	\bibitem{Tycko1985} R. Tycko, A. Pines and J. Guckenheimer, J. Chem. Phys. \textbf{83}, 2775 (1985).
	
	\bibitem{Shaka1984} A. J. Shaka and R. Freeman, J. Magn. Reson. \textbf{59}, 169 (1984).

\bibitem{Ivanov2011} S. S. Ivanov and N. V. Vitanov, Opt. Lett. \textbf{36}, 7 (2011).

\bibitem{Vitanov2011} N. V. Vitanov, Phys. Rev. A \textbf{84}, 065404 (2011).
	%Arbitrarily accurate narrowband composite pulse sequences

\bibitem{Merrill2014} J. T. Merrill, S. C. Doret, G. Vittorini, J. P. Addison, and K. R. Brown, Phys. Rev. A 90, 040301(R) (2014)
% Transformed composite sequences for improved qubit addressing

\bibitem{Wimperis1989} S. Wimperis, J. Magn. Reson. \textbf{83}, 509 (1989).
%Composite pulses with rectangular excitation and inversion profiles,

\bibitem{Kyoseva2013} E. Kyoseva and N. V. Vitanov, Phys. Rev. A \textbf{88}, 063410 (2013).


QINFO %%%%%%%%%%%%%%%%%%%%%%%%%%%%%	

\bibitem{Gulde2003}
S. Gulde, M. Riebe, G. P. T. Lancaster, C. Becher, J. Eschner, H. H\"{a}ffner, F. Schmidt-Kaler, I. L. Chuang and R. Blatt, Nature
\textbf{421}, 48 (2003). % Implementation of the Deutsch–Jozsa algorithm on an ion-trap quantum computer

\bibitem{Schmidt-Kaler2003}
F. Schmidt-Kaler, H. H\"{a}ffner, M. Riebe, S. Gulde, G. P. T. Lancaster, T. Deuschle, C. Becher, C. F. Roos, J. Eschner, and R. Blatt,
Nature \textbf{422}, 408 (2003). %Realization of the Cirac–Zoller controlled-NOT quantum gate

\bibitem{Haffner2008} H. H\"{a}ffner, C. F. Roos, R. Blatt, Phys. Rep. \textbf{469}, 155 (2008).

\bibitem{Timoney2008} N. Timoney, V. Elman, S. Glaser, C. Weiss, M. Johanning, W. Neuhauser, and C. Wunderlich, Phys. Rev. A \textbf{77}, 052334 (2008).

\bibitem{Monz2009} T. Monz, K. Kim, W. H\"{a}nsel, M. Riebe, A. S. Villar, P. Schindler, M. Chwalla, M. Hennrich, and R. Blatt, Phys. Rev. Lett. \textbf{102}, 040501 (2009).


\bibitem{Zarantonello2019}
G. Zarantonello, H. Hahn, J. Morgner, M. Schulte, A. Bautista-Salvador, R. F. Werner, K. Hammerer, and C. Ospelkaus,
Phys. Rev. Lett. \textbf{123}, 260503 (2019)
%Robust and Resource-Efficient Microwave Near-Field Entangling 9Be+ Gate


\bibitem{Torosov2019variable} B. T.~Torosov and N.V.~Vitanov, Phys. Rev. A \textbf{99}, 013402 (2019).
% Arbitrarily accurate variable rotations on the Bloch sphere by composite pulse sequences

\bibitem{Tycko1985jmr} R. Tycko, H. M. Cho, E. Schenider, and A. Pines, J. Magn. Res. \textbf{61}, 90 (1985).


%SCROFULOUS
\bibitem{Cummins2003}
H. K. Cummins, G. Llewellyn, and J. A. Jones, Phys. Rev. A \textbf{67}, 042308 (2003).
%Tackling systematic errors in quantum logic gates with composite rotations

\bibitem{Odedra2012a}
S. Odedra and S. Wimperis, J. Magn. Reson. \textbf{214}, 68 (2012).
%Use of Composite Refocusing Pulses to Form Spin Echoes,

\bibitem{Odedra2012b}
S. Odedra and S. Wimperis, J. Magn. Reson. \textbf{221}, 41 (2012).
%Improved Background Suppression in 1H MAS NMR using Composite Pulses,

\bibitem{Odedra2012c}
S. Odedra, M. J. Thrippleton and S. Wimperis, J. Magn. Reson. \textbf{225}, 81 (2012).
% Dual-Compensated Antisymmetric Composite Refocusing Pulses for NMR,



\bibitem{Jones2013pra}
J. A. Jones, Phys. Rev. A \textbf{87}, 052317 (2013).
%Designing short robust not gates for quantum computation

\bibitem{Jones2013pla}
J. A. Jones, Phys. Lett. A \textbf{377}, 2860 (2013).
%Nested composite NOT gates for quantum computation

\bibitem{Husain2013}
S. Husain, M. Kawamura, J. A. Jones, J. Magn. Res. \textbf{230}, 145 (2013).
%Further analysis of some symmetric and antisymmetric composite pulses for tackling pulse strength errors


\bibitem{Torosov2019phases} B.~T.~Torosov and N.~V.~Vitanov, Phys. Rev. A \textbf{100}, 023410 (2019).
%Composite pulses with errant phases








%%%%%%%%%%%%%%%%%%%%%%%%%%%%%	

	
\end{thebibliography}
\end{document}